\documentclass[referee]{an} 
\usepackage{graphicx}
\usepackage{subeqnarrayart}
\usepackage{times}

\Volume{322}
\Year{2004}
\Month{1}
\Pagespan{223}{227}
\begin{document}
%
%
%
\lhead[\thepage]{R. Caimmi and C. Marmo: homeoidally striated
density profiles}
\rhead[Astron. Nachr./AN~{\bf XXX} (2004) X]{\thepage}
\headnote{Astron. Nachr./AN {\bf 32X} {2004} X, XXX-XXX}
\newenvironment{lefteqnarray}{\arraycolsep=0pt\begin{eqnarray}}
{\end{eqnarray}\protect\aftergroup\ignorespaces}
\newenvironment{lefteqnarray*}{\arraycolsep=0pt\begin{eqnarray*}}
{\end{eqnarray*}\protect\aftergroup\ignorespaces}
\newenvironment{leftsubeqnarray}{\arraycolsep=0pt\begin{subeqnarray}}
{\end{subeqnarray}\protect\aftergroup\ignorespaces}
\newcommand{\displayfrac}[2]{\frac{\displaystyle #1}{\displaystyle #2}}
\newcommand{\diff}{{\rm\,d}}
\newcommand{\appleq}{\stackrel{<}{\sim}}
\newcommand{\appgeq}{\stackrel{>}{\sim}}
\newcommand{\Int}{\mathop{\rm Int}\nolimits}
\newcommand{\Nint}{\mathop{\rm Nint}\nolimits}
\newcommand{\Min}{\mathop{\rm min}\nolimits}
\newcommand{\arcsinh}{\mathop{\rm arcsinh}\nolimits}

\title{Homeoidally striated density profiles: \\
       sequences of virial equilibrium configurations \\
       with constant anisotropy parameters}


\authorrunning{R. Caimmi and C. Marmo}
\titlerunning{homeoidally striated density profiles}

\author{R. Caimmi \and C. Marmo}


\institute{Dipartimento di Astronomia, Universit\`a di Padova,  \\
              Vicolo Osservatorio 2, I-35122 Padova, Italy\\
              \email{caimmi@pd.astro.it}
              }

\date{Received..............................................$\qquad$ Accepted...................................}

\abstract{
%
The formulation of the tensor virial
equations is generalized to unrelaxed
configurations, where virial equilibrium
does not coincide with dynamical (or
hydrostatic) equilibrium.   Special
classes of homeoidally striated ellipsoids
are shown to exhibit similar
properties as Jacobi ellipsoids, and
reduce to Jacobi ellipsoids in the
limiting situation of homogeneous
matter distribution, rigid rotation,
and isotropic residual (i.e. other than
systematic rotation) velocity
distribution.   Accordingly, the
above mentioned density profiles
are defined as homeoidally
striated, Jacobi ellipsoids.
Further investigation is
devoted to the generation of sequences
of virial equilibrium configurations.
Sequences with constant
anisotropy parameters
are studied with more
detail, including both flattened and
elongated, triaxial (in particular, 
both oblate and prolate, axisymmetric)
configurations, and the determination
of the related bifurcation points.
The explicit expression of a number 
of rotation parameters, used in
literature, and the ratio of rms
rotational to rms residual velocity,
is also calculated.   
An application is
made to dark matter haloes
hosting giant galaxies ($M\approx10^
{12}{\rm m}_\odot$),
with regard to assigned initial and
final configuration, following and
generalizing to many respects a procedure
conceived by Thuan \& Gott (1975).
The dependence of the limiting axis ratios, 
below which no configuration is allowed for
the sequence under consideration, on the 
change in mass, total energy, and angular 
momentum, during the evolution, is
illustrated in some representative
situations.
The dependence of the axis ratios, 
$\epsilon_{31}$ and $\epsilon_{21}$, on
a parameter, related to the initial conditions
of the density perturbation, is 
analysed in connection with a few special cases.
The same is done 
for the rotation parameters and the ratio
of rms rotational to rms residual velocity.
Within the range of the rotation parameter,
$\exp_{10}(\log\lambda\mp\sigma_{\log\lambda})=
\exp_{10}(-1.3761\mp0.500)$, consistent with
high-resolution numerical simulations,
the shape of dark matter haloes is mainly
decided by the amount of anisotropy
in residual velocity 
distribution (the above formula uses the
general notation, $\exp_a(x)=a^x$, where the
special case, $a={\rm e}$, reduces to the standard
notation). On the other hand, the
contribution of rotation has only a minor
effect on the meridional plane, and no
effect on the equatorial plane, as bifurcation
points occur for larger values of $\lambda$.
To this respect, dark matter haloes are found 
to resemble giant elliptical galaxies.
   }

   \maketitle
%

\section{Introduction}\label{intro}
Homeoidally striated density profiles make
a useful tool in the description of 
self-gravitating fluids on many respects.
First, the related potential-energy tensors
are expressed as the product between a profile
factor and a shape factor.    Accordingly, a
physical parameter which depends on the ratio
between different components of a same
potential-energy tensor, is a function of the 
axis ratios only (Roberts 1962).   In addition,
the constraints for homeoidally striated
ellipsoids, in connection with both equilibrium
and stability, are left unchanged with respect
to the special case of homogeneous configurations
with same shape (e.g., Binney 1978; Pacheco et 
al. 1989).    Second, a criterion for an upper
limit to the point of bifurcation from
axisymmetric to triaxial configurations, is
formally identical for collisional and
collisionless fluids (Wiegandt 1982a).    On
the other hand, a condition for dynamical
stability depends on the degree of anisotropy
of peculiar velocity distribution (Wiegandt 
1982b).    In this framework, additional features
may be included, such as vorticity and streaming 
motions (e.g., Pacheco et al. 1986, 1989;
Busarello et al. 1989, 1990; and further
references therein).

For the purpose of the current attempt, our
attention shall be restricted to the simplest 
case where both vorticity and streaming motions
are not explicitly investigated, and only 
the contribution from systematic rotation is 
considered separately.
Owing to a generalization of Dedekind's theorem,
to each configuration with no vorticity and
assigned angular velocity, $\Omega$, will
correspond an  ``adjoint'' configuration with
same boundary and distribution of peculiar 
velocity, but with no rotation and vorticity
parameter, $Z=\Omega$ (Pacheco et al. 1989).
Accordingly, to each transition from an initial
to a final, (at least rigidly) rotating
configuration, will correspond a transition
from an initial to a final, vorticating,
adjoint configuration.

The choice of homeoidally striated density 
profiles as a valid approximation to the 
description of self-gravitating fluids, is
motivated by two main reasons.   {\it In
primis}, most astronomical bodies exhibit
ellipsoidal-like shapes and isopycnic surfaces.
{\it Secundo}, homeoidally striated ellipsoids
approximate to a first extent collisional and
collisionless, self-gravitating fluids (e.g.,
Vandervoort 1980; Vandervoort \& Welty 1981; 
Lai et al. 1993), at least with regard to
global properties, involving averages on
the whole mass distribution.   On the other
hand, some caution must be used in dealing
with local properties, involving values
{\it in situ}.    For instance, velocity
dispersion values measured in the central 
region of a galaxy, can be significantly
different for structurally identical
systems (and so characterized by identical,
global velocity dispersions), due to different
orbital structures.

Though classical investigations have been
restricted to figures of dynamical (or
hydrostatic) equilibrium such
as MacLaurin spheroids and Jacobi ellipsoids
(e.g., Jeans 1929, Chaps.\,VIII-IX; 
Chandrasekhar 1966, Chaps.\,5-8), still the
current attempt may be generalized to
figures of virial equilibrium.
Let us define virial equilibrium as
characterized by the validity of the virial
theorem, and relaxed and unrelaxed
configurations as systems where virial 
equilibrium does and does not coincide,
respectively, with dynamical (or 
hydrostatic) equilibrium.

In the special case of relaxed configurations, the
virial theorem is usually formulated by
setting the sum of the potential energy,
and twice the kinetic energy, equal to
zero.   On the other hand, the above
mentioned relation no longer holds for
unrelaxed configurations, and a more
general formulation of the virial equations
is needed.   To this
aim, our attention shall be limited
to homeoidally striated density
profiles, where the rotational
velocity field is subjected to a
number of necessary restrictions.
Then the related sequences of virial equilibrium
configurations with constant rotation
and anisotropy parameters, can be studied in
detail.

An extension of the formulation of the
virial equations to unrelaxed configurations,
and the investigation of sequences where
both rotation and anisotropy parameters
remain unchanged, make the aim
of the current paper.   As an application
of the theory, a
procedure to establish mutual connection
between physical parameters related to
assigned initial and final state of a
density perturbation, may be generalized
to many respects.

About a quarter of century ago, Thuan \& Gott (1975) idealized
elliptical galaxies as MacLaurin spheroids, resulting from
virialization after cosmological expansion and subsequent 
collapse and relaxation of
their parent density perturbations.   Unexpectively, the
above mentioned model provided an answer to the question,
why there are no elliptical galaxies more flattened than E7.   
The method is based on the knowledge of an initial configuration,
assumed to be at turnaround, and a final configuration,
assumed to be a MacLaurin spheroid.   A generalization to
a somewhat more arbitrary initial and final configuration, 
and to other respects,
is expected to be useful for e.g., comparison with (i) results 
of numerical simulations, and (ii) observations.

More specifically, the problem may be formulated as follows.
Let the physical parameters of a density perturbation be
specified at a selected, initial and final configuration, at
redshift $z_i$ and $z_f$, respectively, provided $z_{rec}\ge
z_i\ge z_f\ge z_{vir}$, where the indices, $rec$ and $vir$, 
denote recombination and virialization, respectively.  
Let the density profile be assigned, and the isopycnic
surfaces, i.e. the surfaces of equal density, be similar
and similarly placed ellipsoids.   Let the rotational
velocity field, and the amount of anisotropy in residual
(in particular, peculiar) velocity distribution, also 
be specified.   Under the above mentioned assumptions, 
determine the shape of the final configuration, and
the related physical parameters.

%

The generalization of Thuan \& Gott (1975) procedure
to a special class of homeoidally striated ellipsoids,
together with an application to dark matter haloes
hosting giant galaxies ($M\approx10^{12}{\rm m}_\odot$),
makes a specific application of the theory outlined above.
It shall be analysed in the present attempt.

The current paper is organized as follows.   The general
theory of homeoidally striated density profiles, and the
results of interest to the aim of this investigation,
are outlined in Sect.\,\ref{gente}.   The properties of
a special class of configurations, defined as
homeoidally striated Jacobi ellipsoids, are studied in
Sect.\,\ref{speca} where, in particular, the explicit
expressions of some rotation parameters, known in 
literature, are derived.   Then sequences of virial 
equilibrium configurations are determined, where
both rotation and anisotropy parameters remain
unchanged.   A generalization of the
procedure followed by Thuan \& Gott (1975) to
homeoidally striated density profiles is performed
in Sect.\,\ref{trans}, and a number of special
situations are also investigated.   Furthermore, an
application is made to dark matter haloes hosting
giant galaxies, and the
results are presented and discussed.   Some
concluding remarks are drawn in Sect.\,\ref{conc},   
and a few arguments are treated with more detail in 
the Appendix.   

\section{General theory}
\label{gente}

A general theory for homeoidally striated density 
profiles has been developed in earlier approaches 
(Roberts 1962; Caimmi 1993a; Caimmi \& Marmo 2003, 
hereafter quoted
as CM03), and an interested reader is addressed
therein for deeper insight.   What is relevant for
the current investigation, shall be mentioned and
further developed here.

The isopycnic surfaces are defined by the following
law:
\begin{leftsubeqnarray}
\slabel{eq:profga}
&& \rho=\rho_0f(\xi)~~;\quad f(1)=1~~;\quad\rho_0=
\rho(1)~~; \\
\slabel{eq:profgb}
&& \xi=\frac r{r_0}~~;\quad0\le\xi\le\Xi~~;\quad
\Xi=\frac R{r_0}~~;
\label{seq:profg}
\end{leftsubeqnarray}
where $\rho_0$, $r_0$ are a scaling radius and
a scaling density, respectively, related to a
reference isopycnic surface, and $\Xi$, $R$,
correspond to the boundary.   By ``radius'' we
intend here the radial coordinate of a generic
point on a generic isopycnic surface.   The radius
changes - in general - passing from one point to
one other, along a fixed isopycnic surface, and
it may coincide - in particular - with a
semiaxis of the related ellipsoid.

Let $\Sigma_1$ and $\Sigma_2$ be generic isopycnic
surfaces belonging to a homeoidally striated
density profile, and $r_1$, $r_2$, generic radii
along the same direction.   It can be shown (e.g.,
CM03) that the ratio, $r_1/r_2$, is independent of
direction.   The above result holds, in particular,
in the special case where an isopycnic surface
coincides with the reference isopycnic surface,
according to Eq.\,(\ref{eq:profgb}).   In other
words the scaled radius, $\xi$, is defined as the
ratio between the radius of the isopycnic surface
under consideration, along a generic direction,
and the radius of the reference isopycnic surface,
along the same direction.

The mass, the inertia tensor, and the self-energy
tensor, are:
\begin{lefteqnarray}
\label{eq:M}
&& M=\nu_{mas}M_0~~; \\
\label{eq:Ipq}
&& I_{pq}=\delta_{pq}\nu_{inr}Ma_p^2~~; \\
\label{eq:Espq}
&& (E_{sel})_{pq}=-\frac{GM^2}{a_1}\nu_{sel}
(B_{sel})_{pq}~~; \\
\label{eq:Bspq}
&& (B_{sel})_{pq}=\delta_{pq}\epsilon_{p2}
\epsilon_{p3}A_p~~;\quad B_{sel}=\sum_{s=1}^3
\epsilon_{s2}\epsilon_{s3}A_s~~;
\end{lefteqnarray}
where $\delta_{pq}$ is the Kronecker symbol;
$G$ is the constant of gravitation; $\nu_
{mas}$, $\nu_{inr}$, $\nu_{sel}$, are profile
factors i.e. depend only on the density profile
via the scaled radius, $\Xi$; $a_1$, $a_2$,
$a_3$, are semiaxes; $\epsilon_{pq}=a_p/a_q$ 
are axis
ratios; $A_1$, $A_2$, $A_3$, are shape factors
i.e. depend only on the axis ratios; and $M_0$
is the mass of a homogeneous ellipsoid with
same density and boundary as the reference
isopycnic surface:
\begin{equation}
\label{eq:M0}
M_0=\frac{4\pi}3\rho_0a_{01}a_{02}a_{03}~~;
\end{equation}
where $a_{01}$, $a_{02}$, $a_{03}$, are the
semiaxes of the ellipsoid bounded by the 
reference isopycnic surface.
Owing to a different normalization of the
physical quantities, the profile parameters,
$\nu$, have a different expression with
respect to CM03.   For further details, see 
Appendix A.   Values of shape factors
related to limiting configurations (round,
oblate, prolate, flat, and oblong) are derived
in Appendix B.

In dealing with angular momentum and rotational
energy, the preservation of (triaxial) ellipsoidal
shape imposes severe constraints on the rotational
velocity field.    Leaving an exhaustive investigation
to more refined approaches, our attention shall
be limited here to a restricted number of special cases,
namely: (i) rigid rotation about a principal axis,
and (ii) differential rotation about a symmetry
axis.   By ``rotation'' it is intended, of course,
circular rotation.

In the special case of rigid rotation about a
principal axis, let it be $x_3$, the
angular-momentum vector and the rotational-energy
tensor are:
\begin{lefteqnarray}
\label{eq:Jpqrr}
&& J_s=\delta_{s3}(I_{pp}+I_{qq})\Omega_s=
\delta_{s3}\nu_{inr}Ma_p(1+\epsilon_{qp}^2)
(v_{rot})_p~~; \quad p\ne q\ne s~~; \\
\label{eq:Erpqrr}
&& (E_{rot})_{pq}=\frac12I_{pq}\delta_{s3}
\Omega_s^2=\frac12\delta_{s3}\delta_{pq}(1-
\delta_{ps})\nu_{inr}Ma_p^2\Omega^2 \nonumber \\
&& \phantom{(E_{rot})_{pq}}=\frac12
\delta_{s3}\delta_{pq}(1-\delta_{ps})\nu_{inr}
M\left[(v_{rot})_p\right]^2~~;\qquad p\ne s~~;
\qquad q\ne s~~;
\end{lefteqnarray}
where $\Omega$ is the angular velocity related to
the axis, $x_3$, and $(v_{rot})_p$ is the rotational
velocity at the end of the semiaxis, $a_p$, $p=1,2$.

The module of the above vector and the trace of the
above tensor, which make the related angular momentum 
and rotational energy, respectively, read:
\begin{lefteqnarray}
\label{eq:Jrr}
&& J=\nu_{inr}Ma_p^2(1+\epsilon_{qp}^2)\Omega=
\nu_{inr}Ma_p(1+\epsilon_{qp}^2)(v_{rot})_p~~; \\
\label{eq:Errr}
&& E_{rot}=\frac12\nu_{inr}Ma_p^2(1+\epsilon_{qp}^2)
\Omega^2=\frac12\nu_{inr}M(1+\epsilon_{qp}^2)\left[
(v_{rot})_p\right]^2~~;
\end{lefteqnarray}
for further details, see CM03.

In the special case of differential rotation about
a symmetry axis, let it be $x_3$, our attention
shall be restricted to rotational (with respect
to $x_3$ axis) velocity distributions which obey 
the law:
\begin{equation}
\label{eq:rvrot}
\frac{v_{rot}(r,\theta)}{v_{rot}(R,\theta)}=\frac
{v_{rot}(a_p^\prime,0)}{v_{rot}(a_p,0)}~~;
\end{equation}
or equivalently, to angular (with respect to
$x_3$ axis) velocity distributions which obey 
the law:
\begin{equation}
\label{eq:rvang}
\frac{\Omega(r,\theta)}{\Omega(R,\theta)}=\frac
{\Omega(a_p^\prime,0)}{\Omega(a_p,0)}~~;
\end{equation}
where $(r,\theta)$, $(R,\theta)$, represent a
point on a generic isopycnic surface and on the
boundary, respectively, along a fixed radial
direction, and $(a_p^\prime,0)$, $(a_p,0)$,
$p=1,2,$ represent
the end of the corresponding equatorial semiaxis.
It is worth noticing that, in particular, either
rotational or angular velocity are allowed to be
constant everywhere.

The equivalence of Eqs.\,(\ref{eq:rvrot}) and
(\ref{eq:rvang}) is owing to the similarity
between isopycnic surfaces, which translates
into the equivalent relations:
\begin{equation}
\label{eq:rrR}
\frac rR=\frac{a^\prime}a~~;\quad\frac r{r_0}
=\frac{a^\prime}{a_0}~~;
\end{equation}
as outlined above in connection with ellipsoidal
configurations.

The angular-momentum vector and the rotational-energy
tensor are:
\begin{lefteqnarray}
\label{eq:Jpqrd}
&& J_s=2\delta_{s3}\eta_{anm}\nu_{anm}Mav_{rot}~~; \\
\label{eq:Erpqrd}
&& (E_{rot})_{pq}=\delta_{pq}(1-\delta_{p3})\eta_{rot}
\nu_{rot}Mv_{rot}^2~~;
\end{lefteqnarray}
where $v_{rot}$ is the rotational velocity at the end
of the equatorial semiaxis, $a$; $\eta_{anm}$,
$\eta_{rot}$, depend on the boundary and on the
distribution of angular velocity therein, and then
may be conceived as shape factors; on the other hand,
$\nu_{anm}$, $\nu_{rot}$, depend on the density
profile and on the radial distribution of angular 
velocity, and then are genuine profile factors.

The module of the above vector and the trace of
the above tensor, which make the
related angular momentum and rotational energy,
respectively, read:
\begin{lefteqnarray}
\label{eq:Jrd}
&& J=2\eta_{anm}\nu_{anm}Mav_{rot}~~; \\
\label{eq:Errd}
&& E_{rot}=2\eta_{rot}\nu_{rot}Mv_{rot}^2~~;
\end{lefteqnarray}
for further details, see CM03.

To save space, let us merge Eqs.\,(\ref
{eq:Jpqrr}) and (\ref{eq:Jpqrd}); (\ref
{eq:Erpqrr}) and (\ref{eq:Erpqrd}); as:
\begin{lefteqnarray}
\label{eq:Jpq}
&& J_s=\delta_{s3}\eta_{anm}\nu_{anm}M
a_p(1+\epsilon_{qp}^2)(v_{rot})_p~~;
\quad p\ne q\ne s~~; \\
\label{eq:Erpq}
&& (E_{rot})_{pq}=\delta_{pq}(1-\delta_{p3})
\eta_{rot}\nu_{rot}M\left[(v_{rot})_p\right]^2~~;
\end{lefteqnarray}
and Eqs.\,(\ref
{eq:Jrr}) and (\ref{eq:Jrd}); (\ref
{eq:Errr}) and (\ref{eq:Errd}); as:
\begin{lefteqnarray}
\label{eq:J}
&& J=\eta_{anm}\nu_{anm}M(1+\epsilon_{21}^2)
a_1(v_{rot})_1~~; \\
\label{eq:Er}
&& E_{rot}=\eta_{rot}\nu_{rot}M(1+\epsilon_
{21}^2)\left[(v_{rot})_1\right]^2~~;
\end{lefteqnarray}
respectively, where $\eta_{anm}\nu_{anm}=
\nu_{inr}$ and $\eta_{rot}\nu_{rot}=\nu_
{inr}/2$ in the special case of rigid rotation;
for further details, see CM03.

An expression of the rotational energy, $E_
{rot}$, as a function of the angular momentum,
$J$, is obtained by the combination of 
Eqs.\,(\ref{eq:J}) and (\ref{eq:Er}), as:
\begin{leftsubeqnarray}
\slabel{eq:ErJa}
&& E_{rot}=\frac {J^2}{Ma_1^2}\nu_{ram}B_{ram}~~; \\
\slabel{eq:ErJb}
&& \nu_{ram}=\frac{\nu_{rot}}{\nu_{anm}^2}~~; \\
\slabel{eq:ErJc}
&& (B_{ram})_{pq}=\delta_{pq}(1-\delta_{p3})\frac
{\eta_{rot}}{\eta_{anm}^2}\frac{\epsilon_{p1}^2}
{(1+\epsilon_{21}^2)^2}~~; \\
\slabel{eq:ErJd}
&& B_{ram}=\frac{\eta_{rot}}{\eta_{anm}^2}\frac1
{1+\epsilon_{21}^2}~~;
\label{seq:ErJ}
\end{leftsubeqnarray}
and the ratio of rotational to self-potential
energy, taken with the positive sign, ${\cal
E}_{rot}$, is obtained by the combination of
Eqs.\,(\ref{eq:Espq}), extended to the trace
of the related tensor, and (\ref{seq:ErJ}), as:
\begin{leftsubeqnarray}
\slabel{eq:ErEa}
&& {\cal E}_{rot}=-\frac{E_{rot}}{E_{sel}}=\frac
{\nu_{ram}}{\nu_{sel}}\frac{B_{ram}}{B_{sel}}h~~; \\
\slabel{eq:ErEb}
&& h=\frac{J^2}{GM^3a_1}~~;
\label{seq:ErE}
\end{leftsubeqnarray}
where both ${\cal E}_{rot}$ (e.g., Ostriker \&
Peebles 1973) and $h$ (e.g., Caimmi 1980) may 
be conceived as rotation parameters. 

At this stage, we have all the ingredients
for generalizing the usual formulation of
the tensor virial equations (e.g., Chandrasekhar
1961, Chap.\,13, \S 117; Brosche 1970; Brosche
et al. 1983), to include  
unrelaxed configurations, in particular
homeoidally striated density profiles.

\section{The tensor virial equations for
homeoidally striated density profiles}
\label{speca}
\subsection{Basic theory}
\label{bathe}

As outlined in Sect.\,1,
homeoidally striated density profiles make
a useful tool in the description of 
self-gravitating fluids on many respects.
Though classical investigations have been
restricted to figures of dynamical (or
hydrostatic) equilibrium such
as MacLaurin spheroids and Jacobi ellipsoids
(e.g., Jeans 1929, Chaps.\,VIII-IX; 
Chandrasekhar 1966, Chaps.\,5-8), still the
current attempt may be generalized to
figures of virial equilibrium.   To this
respect, it is worth remembering that 
the virial theorem holds for values of 
parameters averaged over a sufficiently 
long time, $t\gg T$, where $T$ is a 
characteristic period of the system (e.g.,
Landau \& Lifchitz 1966, Chap.\,II, \S\,10).

Let us define virial equilibrium as
characterized by the validity of the virial
theorem, and relaxed and unrelaxed
configurations as systems where virial 
equilibrium does and does not coincide,
respectively, with dynamical (or 
hydrostatic) equilibrium.  For instance,
a compressible homogeneous sphere undergoing
coherent oscillations, is in virial equilibrium:
then the generic configuration is unrelaxed, even
if a special one exists, equal in shape to its
relaxed counterpart.

In the special case of relaxed configurations, the
virial theorem is usually formulated by
setting the sum of the potential energy,
and twice the kinetic energy, equal to
zero.   On the other hand, the above
mentioned relation no longer holds for
unrelaxed configurations, and a more
general formulation of the virial equations
is needed.   To this
aim, our attention shall be limited
to homeoidally striated density
profiles, where the rotational
velocity field is subjected to the
restrictions mentioned in Sect.\,\ref
{gente}.

With regard to unrelaxed configurations,
let us define the residual kinetic-energy
tensor, $(E_{res})_{pq}$, as owing to any
motion other than systematic rotation 
e.g., random motions, streaming motions, 
radial motions; and the effective,
residual kinetic-energy tensor, $(\tilde
{E}_{res})_{pq}$, as the right amount 
needed for the configuration of interest
to be relaxed i.e. the validity of the 
tensor virial equations:
\begin{equation}
\label{eq:virte}
(E_{sel})_{pq}+2(E_{rot})_{pq}+2
(\tilde{E}_{res})_{pq}=0~~;
\end{equation}
then the components of the effective
anisotropy tensor, defined as:
\begin{leftsubeqnarray}
\label{eq:ziefa}
&& \tilde{\zeta}_{pq}=\frac{(\tilde{E}_{res})_{pq}}
{\tilde{E}_{res}}~~;\qquad p=1,2,3~~;\qquad
q=1,2,3~~; \\
\label{eq:ziefb}
&&\sum_{p=1}^3\tilde{\zeta}_{pp}=1~~;\qquad
0\le\tilde{\zeta}_{pp}\le1~~;\qquad
\tilde{\zeta}_{pq}=0~~;\qquad p\ne q~~;
\label{seq:zief}
\end{leftsubeqnarray}
may be conceived as effective anisotropy
parameters (with respect to the effective 
residual kinetic-energy tensor).
Similarly, the components of the generalized
anisotropy tensor, defined as:
\begin{leftsubeqnarray}
\label{eq:zirfa}
&& \zeta_{pq}=\frac{(\tilde{E}_{res})_{pq}}
{E_{res}}~~;\qquad p=1,2,3~~;\qquad q=1,2,3~~; \\
\label{eq:zirfb}
&& \sum_{p=1}^3
\zeta_{pp}=\frac{\tilde{E}_{res}}{E_{res}}=
\zeta~~;\qquad0\le\zeta_{pp}\le\zeta~~;\qquad
\zeta_{pq}=0~~;\qquad p\ne q~~; 
\label{seq:zirf}
\end{leftsubeqnarray}
may be conceived as generalized anisotropy
parameters (with respect to the generalized
residual kinetic-energy tensor).
%

The scalar, $\zeta$, may be thought of as a
virial index, where $\zeta=1$ corresponds
to null virial, $(E_{sel})_{pq}+2(E_{rot})_
{pq}+2(E_{res})_{pq}=0$ (which doe
not necessarily imply a relaxed configuration),
$\zeta>1$
to negative virial, and $\zeta<1$ to positive virial.

The combination of Eqs.\,(\ref{seq:zief}) and (\ref
{seq:zirf}) yields:
\begin{equation}
\label{eq:zaze}
\tilde{\zeta}_{pp}=\frac{\zeta_{pp}}\zeta~~;\qquad
p=1,2,3~~;
\end{equation}
and the specification of the effective and
generalized anisotropy parameters,
$\tilde{\zeta}_{pp}$ and $\zeta_{pp}$, and the
residual kinetic energy, $E_{res}$, implies
the specification of the effective, residual
kinetic-energy tensor, $(\tilde{E}_{res})_
{pq}$.

In the special case of relaxed configurations,
the effective, residual and actual kinetic-energy
tensors coincide, $(\tilde{E}_{res})_{pq}
=(E_{res})_{pq}$, i.e. $\zeta=1$.   Then the
limiting situation of relaxed configurations
may directly be obtained from the results of
the current Section, in the limit $\zeta\to1$.

Using Eqs.\,(\ref{eq:Espq}), (\ref{eq:Bspq}), 
(\ref{eq:Erpq}), (\ref{seq:ErJ}), and (\ref
{seq:zirf}), Eq.\,(\ref{eq:virte}) takes the 
more explicit expression:
\begin{equation}
\label{eq:vires}
-\nu_{sel}\frac{GM^2}{a_1}(B_{sel})_{pp}+
2\nu_{ram}\frac{J^2}{Ma_1^2}(B_{ram})_{pp}+2
\zeta_{pp}E_{res}=0~~;
\end{equation}
leaving aside the trivial case of non-diagonal
terms, where they reduce to identities, $0=0$.
The special choice, $p=3$, reads: 
\begin{equation}
\label{eq:virt3}
-\nu_{sel}\frac{GM^2}{a_1}(B_{sel})_{33}+2
\zeta_{33}E_{res}=0~~;
\end{equation}
which allows the following expression of
the residual kinetic energy:
\begin{equation}
\label{eq:Epec}
E_{res}=\frac12\frac{\nu_{sel}}{\zeta_{33}}
\frac{GM^2}{a_1}(B_{sel})_{33}~~;
\end{equation}
as first outlined in dealing with MacLaurin
spheroids (Brosche 1970).

The combination of Eqs.\,(\ref{eq:vires}) and
(\ref{eq:Epec}) yields:
\begin{equation}
\label{eq:virt1}
\nu_{sel}\frac{GM^2}{a_1}\left[(B_{sel})_{qq}-
\frac{\zeta_{qq}}{\zeta_{33}}(B_{sel})_{33}\right]-
2\nu_{ram}\frac{J^2}{Ma_1^2}(B_{ram})_{qq}=0~~;
\quad q=1,2~~;
\end{equation}
leaving aside the trivial case, $p=3$, where
it reduces to an identity, $0=0$.   A necessary
condition for satisfiyng Eq.\,(\ref{eq:virt1}),
or (\ref{eq:virte}), using Eqs.\,(\ref{eq:Bspq}),
may be written as:
\begin{equation}
\label{eq:cone}
\frac{\zeta_{qq}}{\zeta_{33}}\le\frac{(B_{sel})_{qq}}
{(B_{sel})_{33}}=\frac{A_q}{\epsilon_{3q}^2A_3}
~~;\qquad q=1,2~~;
\end{equation}
which is the natural extension of its counterpart
related to axisymmetric, relaxed configurations 
(Wiegandt1982a-b).

\subsection{Rotation parameters}
\label{rotpa}
The amount
of rotation parameter, $h$, needed to satisfy
Eqs.\,(\ref{eq:vires}), can be
determined via Eqs.\,(\ref{eq:ErJb}) and
(\ref{eq:ErEb}) by use of
Eq.\,(\ref{eq:virt1}). The result is:
\begin{equation}
\label{eq:virh}
h=\frac12\frac{\nu_{sel}}{\nu_{ram}}\frac{(B_
{sel})_{qq}}{(B_{ram})_{qq}}\left[1-\frac{\zeta
_{qq}}{\zeta_{33}}\frac{(B_{sel})_{33}}{(B_{sel})_{qq}}
\right]~~;\quad q=1,2~~;
\end{equation}
and the related value of the rotation parameter, 
${\cal E}_{rot}$, owing to Eqs.\,(\ref{eq:ErEa}) and
(\ref{eq:virh}), takes the explicit expression:
\begin{equation}
\label{eq:rErs}
{\cal E}_{rot}=\frac12\frac{B_{ram}}{B_{sel}}
\frac{(B_{sel})_{qq}}{(B_{ram})_{qq}}\left[1-
\frac{\zeta_{qq}}{\zeta_{33}}\frac{(B_{sel})_{33}}
{(B_{sel})_{qq}}\right]~~;\quad q=1,2~~;
\end{equation}
where $\zeta_{qq}=\zeta_{33}=1/3$ for isotropic,
residual velocity distributions (i.e. related
to the residual kinetic energy); $\epsilon_
{q2}=\epsilon_{q1}=1$ for spheroidal
configurations; $(1+\epsilon_{21}^2)B_{ram}
=1/2$, $\nu_{ram}=5$, for rigidly rotating
configurations; $\nu_{sel}=3/10$ for homogeneous
configurations.

The validity of Eq.\,(\ref{eq:virh}), or
equivalently Eq.\,(\ref{eq:rErs}), implies
the following relation:
\begin{equation}
\label{eq:Brszt}
\zeta_{33}\left[(B_{sel})_{11}(B_{ram})_{22}-
(B_{sel})_{22}(B_{ram})_{11}\right]=
(B_{sel})_{33}\left[\zeta_{11}(B_{ram})_{22}-
\zeta_{22}(B_{ram})_{11}\right]~~;
\end{equation}
which links the axis ratios, $\epsilon_
{21}$ and $\epsilon_{31}$, with the
anisotropy parameters, $\zeta_{11}$ and
$\zeta_{22}$, owing to Eq.\,(\ref{seq:zirf}).
For fixed values of the (generalized or
effective) anisotropy parameters, Eq.\,(\ref
{eq:Brszt}) describes the change of an axis
ratio as a function of the other one, along
the related sequence of virial equilibrium
configurations.

The rotation parameters, $h$ and ${\cal E}_{rot}$,
expressed by Eqs.\,(\ref{eq:virh}) and (\ref
{eq:rErs}), respectively, exhibit a substantial
difference: while the former depends on profile
factors, $\nu_{sel}$ and $\nu_{ram}$, the latter
does not.   Then the value of ${\cal E}_{rot}$
remains unchanged for homeoidally striated
ellipsoids with fixed axis ratios and anisotropy
parameters.   With regard to $h$, the special
case of Jacobi ellipsoids reads (e.g., Chandrasekhar
1962, but the notation is different therein):
\begin{equation}
\label{eq:hJ}
h=\frac3{50}\frac{(1+\epsilon_{21}^2)^2}{\epsilon
_{q1}^2}[(B_{sel})_{qq}-(B_{sel})_{33}]~~;\quad 
q=1,2~~;
\end{equation}
and the special case of MacLaurin spheroids
corresponds to $\epsilon_{21}=1$, $(B_{ram})_
{qq}=1/8$, owing to Eq.\,(\ref{eq:ErJc}) and
rigid rotation.   For
further details on MacLaurin spheroids and 
Jacobi ellipsoids see e.g., Jeans (1929, 
Chap.\,VIII); Chandrasekhar (1969, Chaps.\,5-6);
Caimmi (1996a).

At this stage, let us define the rotation parameter:
\begin{equation}
\label{eq:vg}
\upsilon=\frac{\Omega^2}{2\pi G\bar{\rho}}~~;
\end{equation}
where $\Omega=\Omega(a_1,0)$ is the angular
velocity at the end of the major equatorial semiaxis,
denoted as $a_1$, and $\bar{\rho}$ is the mean density
of the ellipsoid:
\begin{equation}
\label{eq:rome}
\bar{\rho}=\frac3{4\pi}\frac M{a_1a_2a_3}~~;
\end{equation}
the above definition of the rotation parameter,
$\upsilon$, makes a generalization of some
special cases mentioned in the literature
(e.g., Jeans 1929, Chap.\,IX, \S 232; 
Chandrasekhar \& Leboviz 1962).   The
combination of Eqs.\,(\ref{eq:J}), (\ref
{eq:ErEb}), (\ref{eq:vg}), and (\ref
{eq:rome}) yields:
\begin{equation}
\label{eq:hups}
h=\frac32\eta_{anm}^2\nu_{anm}^2\frac{(1+\epsilon_{21}^2
)^2}{\epsilon_{21}\epsilon_{31}}\upsilon~~;
\end{equation}
which links the rotation parameters, $h$ and
$\upsilon$.   

The combination of Eqs.\,(\ref{eq:ErJb}), 
(\ref{eq:ErJc}), (\ref{eq:virh}), and 
(\ref{eq:hups}) allows the explicit expression:
\begin{equation}
\label{eq:upsh}
\upsilon=\frac13\frac{\nu_{sel}}{\eta_{rot}\nu_
{rot}}\epsilon_{2q}\epsilon_{3q}(B_{sel})
_{qq}\left[1-\frac{\zeta_{qq}}{\zeta_{33}}\frac
{(B_{sel})_{33}}{(B_{sel})_{qq}}\right]~~;
\quad q=1,2~~;
\end{equation}
which, in the special case of rigidly
rotating, homogeneous configurations
with isotropic peculiar velocity 
distribution, reduces to a known
relation for Jacobi ellipsoids and,
with the additional demand of axial
symmetry, to a known relation for
MacLaurin spheroids (e.g., Jeans
1929, Chap.\,VIII, \S\S 189-193;
Chandrasekhar 1969, Chap.\,5, \S
32, Chap.\,6, \S 39; Caimmi 1996a).

In conclusion, Jacobi ellipsoids
and MacLaurin spheroids may be
conceived as limiting cases within
the class of homeoidally
striated ellipsoids under discussion
i.e. characterized by rotational or angular
velocity distribution expressed by
Eqs.\,(\ref{eq:rvrot}) or (\ref
{eq:rvang}), respectively.   For 
this reason, we define 
homeoidally striated ellipsoids or
spheroids of the kind considered,
as homeoidally striated Jacobi
ellipsoids or MacLaurin spheroids,
respectively.   

On the other hand, 
one must keep in mind that Jacobi
ellipsoids and MacLaurin spheroids
are relaxed configurations i.e. in dynamical
equilibrium (e.g., Jeans 1929,
Chap.\,VIII, \S\S 189-193;
Chandrasekhar 1969, Chap.\,5, \S
32, Chap.\,6, \S39), while their
homeoidally striated counterparts
are unrelaxed configurations i.e. in virial
equilibrium but, at least in general,
not in dynamical (or hydrostatic) 
equilibrium (e.g.,
Tassoul 1978, Chap.\,4, Sect.\,4.3;
Chambat 1994; Caimmi 1996a,b).   In
addition, homeoidally striated
ellipsoids approximate to a first
extent self-gravitating fluids, at
least in connection with typical
values of physical parameters,
averaged over the whole volume
(e.g., Vandervoort 1980; Vandervoort
\& Welty 1981; Lai et al. 1993).

\subsection{Isotropic and anisotropic
configurations}\label{isanc}

The ratios between anisotropy parameters
can be obtained via Eq.\,(\ref{eq:virh}).
The result is:
\begin{lefteqnarray}
\label{eq:zq3}
&& \frac{\zeta_{qq}}{\zeta_{33}}=\frac{(B_{sel})_{qq}}
{(B_{sel})_{33}}\left[1-2h\frac{\nu_{ram}}{\nu_
{sel}}\frac{(B_{ram})_{qq}}{(B_{sel})_{qq}}
\right]~~;\quad q=1,2~~; \\
\label{eq:z12}
&& \frac{\zeta_{11}}{\zeta_{22}}=\frac{\nu_{sel}(B_
{sel})_{11}-2h\nu_{ram}(B_{ram})_{11}}{\nu_
{sel}(B_{sel})_{22}-2h\nu_{ram}(B_{ram})_
{22}}~~;
\end{lefteqnarray}
and the combination of Eqs.\,(\ref{seq:zirf}) 
and (\ref{eq:zq3}) yields:
\begin{equation}
\label{eq:z3}
\frac{\zeta_{33}}{\zeta}=\frac{(B_{sel})_{33}}
{B_{sel}}\left[1-2h\frac{\nu_{ram}}{\nu_{sel}}
\frac{B_{ram}}{B_{sel}}\right]^{-1}~~;
\end{equation}
which provides an alternative expression to
Eqs.\,(\ref{eq:virh}), (\ref{eq:rErs}), and
(\ref{eq:upsh}), as:
\begin{lefteqnarray}
\label{eq:hz}
&& h=\frac12\frac{\nu_{sel}}{\nu_{ram}}\frac{B_{sel}}
{B_{ram}}\left[1-\frac{\zeta}{\zeta_{33}}\frac{(B_{sel})_
{33}}{B_{sel}}\right]~~; \\
\label{eq:Erz}
&& {\cal E}_{rot}=\frac12\left[1-\frac{\zeta}{\zeta_{33}}
\frac{(B_{sel})_{33}}{B_{sel}}\right]~~; \\
\label{eq:upz}
&& \upsilon=\frac13\frac{\nu_{sel}}{\eta_{rot}\nu_
{rot}}\frac{\epsilon_{21}\epsilon_{31}}{1+\epsilon_
{21}^2}B_{sel}\left[1-\frac{\zeta}{\zeta_{33}}\frac
{(B_{sel})_{33}}{B_{sel}}\right]~~;
\end{lefteqnarray}
where the rotation parameter, $h$, is expressed
as a function of profile parameters, shape
parameters, and the effective anisotropy parameter, 
$\tilde{\zeta}_{33}=\zeta_{33}/\zeta$.   In addition, 
Eqs.\,(\ref{eq:hz}), (\ref{eq:Erz}), and (\ref
{eq:upz}), have the same formal expression as in 
the special case of axisymmetric configurations 
(e.g., Caimmi 1996b).         

The combination of Eqs.\,(\ref{eq:zq3}) and (\ref
{eq:z3}) yields:
\begin{equation}
\label{eq:zq}
\frac{\zeta_{pp}}{\zeta}=\frac{(B_{sel})_{pp}}{B_{sel}}
\left[1-2h\frac{\nu_{ram}}{\nu_{sel}}\frac{(B_{ram})
_{pp}}{(B_{sel})_{pp}}\right]\left[1-2h\frac{\nu_{ram}}
{\nu_{sel}}\frac{B_{ram}}{B_{sel}}\right]^{-1}~~;\quad 
p=1,2,3~~;
\end{equation}
in the limiting case of a vanishing rotation,
$h\to0$, Eqs.\,(\ref{eq:zq}) reduce to:
\begin{equation}
\label{eq:zp}
\frac{\zeta_{pp}}\zeta=\frac{(B_{sel})_{pp}}
{B_{sel}}~~;\quad p=1,2,3~~;
\end{equation}
where flattening and/or elongation are owing
to anisotropic, residual velocity distribution
(i.e. related to the residual kinetic energy)
only.

In the special case of isotropic configurations,
$\zeta_{pp}=1/3$, $p=1,2,3$, Eq.\,(\ref{eq:hz})
reduces to:
\begin{equation}
\label{eq:hi}
h_{iso}=\frac12\frac{\nu_{sel}}{\nu_{ram}}\frac
{B_{sel}}{B_{ram}}\left[1-3\frac{(B_{sel})_
{33}}{B_{sel}}\right]~~;
\end{equation}
and an equivalent expression may be obtained
from Eq.\,(\ref{eq:virh}).   The combination 
of Eqs.\,(\ref{eq:hz}), and (\ref{eq:hi}) 
yields:
\begin{equation}
\label{eq:hhi}
h=h_{iso}+\frac12\frac{\nu_{sel}}{\nu_{ram}}\frac
{3\tilde{\zeta}_{33}-1}{\tilde{\zeta}_{33}}\frac{(B_
{sel})_{33}}{B_{ram}}~~;
\end{equation}
which shows the effect of polar (with respect to
equatorial) anisotropy for a fixed shape.   

Values $\tilde{\zeta}_{33}<1/3$ make $h<h_{iso}$ i.e. 
less rotation
is needed for a fixed flattening, as a contribution
from anisotropy arises.    The anisotropy
parameter cannot be lower than a critical value,
which corresponds to $h=0$; if otherwise, the
shape under consideration would not be related
to a configuration of virial equilibrium.   
On the contrary, values $\tilde{\zeta}_{33}>1/3$ make 
$h>h_{iso}$ i.e. more rotation is needed for a fixed 
flattening, as a contribution from anisotropy 
arises, but in the opposite sense with respect
to the former case.   

Following a similar procedure, Eq.\,(\ref
{eq:upz}) reads:
\begin{leftsubeqnarray}
\slabel{eq:via}
&& \upsilon=\upsilon_{iso}+\frac13\frac{\nu_{sel}}
{\eta_{rot}\nu_{rot}}\frac{\epsilon_{21}\epsilon_{31}}
{1+\epsilon_{21}^2}\frac{3\tilde{\zeta}_{33}-1}{\tilde
{\zeta}_3}(B_{sel})_{33}~~; \\
\slabel{eq:vib}
&& \upsilon_{iso}=\frac13\frac{\nu_{sel}}
{\eta_{rot}\nu_{rot}}\frac{\epsilon_{21}\epsilon_{31}}
{1+\epsilon_{21}^2}\left[B_{sel}-3(B_{sel})_{33}\right]~~;
\label{seq:vi}
\end{leftsubeqnarray}
where the effect of anisotropy occurs as discussed
above.    

Owing to Eqs.\,(\ref{eq:Espq}), (\ref{eq:virte}), 
and (\ref{seq:zirf}), the total energy of a 
homeoidally striated, Jacobi ellipsoid is:
\begin{equation}
\label{eq:E}
E=\frac12E_{sel}+(1-\zeta)E_{res}=-\frac12\nu_
{sel}\frac{GM^2}{a_1}B_{sel}\left[1-\frac{1-
\zeta}\zeta\frac\zeta{\zeta_{33}}\frac{(B_{sel})_
{33}}{B_{sel}}\right]~~;
\end{equation}
and the combination of Eqs.\,(\ref{eq:ErEb}), 
(\ref{eq:virh}), and (\ref{eq:E}) yields the
explicit expression of the rotation parameter,
$\lambda$ (e.g., Peebles 1969):
\begin{lefteqnarray}
\label{eq:lam2}
&& \lambda^2=-\frac{J^2E}{G^2M^5}=\frac14\frac
{\nu_{sel}^2}{\nu_{ram}}\frac{(B_{sel})_{qq}}
{(B_{ram})_{qq}}\left[1-\frac{\zeta_{qq}}{\zeta_{33}}
\frac{(B_{sel})_{33}}{(B_{sel})_{qq}}\right]
B_{sel}\left[1-\frac{1-
\zeta}\zeta\frac\zeta{\zeta_{33}}\frac{(B_{sel})_
{33}}{B_{sel}}\right]~~;\nonumber \\ 
&& q=1,2~~;
\end{lefteqnarray}
an equivalent expression may be obtained using
Eq.\,(\ref{eq:hz}) instead of (\ref{eq:virh});
the result is:
\begin{lefteqnarray}
\label{eq:lam3}
&& \lambda^2=\frac14\frac
{\nu_{sel}^2}{\nu_{ram}}\frac{B_{sel}^2}
{B_{ram}}\left[1-\frac\zeta{\zeta_{33}}
\frac{(B_{sel})_{33}}{B_{sel}}\right]
\left[1-\frac{1-
\zeta}\zeta\frac\zeta{\zeta_{33}}\frac{(B_{sel})_
{33}}{B_{sel}}\right]=\nonumber \\
&& \phantom{\lambda^2}=\frac12\nu_{sel}B_{sel}
h\left[1-\frac{1-\zeta}\zeta\frac\zeta{\zeta_{33}}
\frac{(B_{sel})_{33}}{B_{sel}}\right]~~;
\end{lefteqnarray}
which is a generalization of an expression, 
previously calculated for homeoidally striated
Jacobi ellipsoids, where $\zeta_{11}=\zeta_{22}$
(Caimmi 1993b).   

Let us define, using Eq.\,(\ref{eq:Epec}),
a typical rms residual velocity, $<v_
{res}^2>=2E_{res}/M$, as:
\begin{equation}
\label{eq:vpec}
<v_{res}^2>=\frac{\nu_{sel}}{\zeta_{33}}\frac
{GM}{a_1}{(B_{sel})_{33}}~~;
\end{equation}
and define, using Eqs.\,(\ref{eq:ErJa})
and (\ref{eq:virt1}), a 
typical square rotational velocity, $<v_
{rot}^2>=2E_{rot}/M$, as:
\begin{equation}
\label{eq:vrot}
<v_{rot}^2>=\nu_{sel}\frac{GM}{a_1}\left[
B_{sel}-\frac\zeta{\zeta_{33}}(B_{sel})_{33}\right]~~;
\end{equation}
then the combination of Eqs.\,(\ref{eq:vpec})
and (\ref{eq:vrot}) yields:
\begin{equation}
\label{eq:vrope}
\chi_v^2=\frac{<v_{rot}^2>}{<v_{res}^2>}=\zeta
\left[\frac{\zeta_{33}}\zeta\frac{B_{sel}}{(B_{sel})_
{33}}-1\right]~~;
\end{equation}
which for anisotropic, spheroidal configurations
rotating around their symmetry axis, attains a
different expression with respect to e.g.,
Binney \& Tremaine (1987, Chap.\,4, \S\,3),
owing to a different definition of $<v_{res}^2>$
therein.    The rms velocity ratio, $\chi_v$,
may also be considered as a rotation parameter.

However, some caution has to be used in the
interpretation of these results.   According
to Eq.\,(\ref{eq:vrope}) the rms velocity ratio,
$\chi_v$, depends only on the axis ratios, 
$\epsilon_{21}$ and $\epsilon_{31}$, and the 
generalized anisotropy 
parameter, $\zeta_{33}$ (e.g., Binney 1976;
Binney \& Tremaine 1987, Chap.\,4, \S\,3).
Then homeoidally striated ellipsoids with
different density profiles and same generalized
anisotropy parameter and axis ratios, have the 
same rms
velocity ratio.   On the other hand, it is
not true for self-gravitating fluids, at
least in the limit of isotropic, peculiar
velocity distributions, where the rms velocity
ratio tends to zero as the density profile
tends to the Roche limit, i.e. a mass point 
surrounded by a massless atmosphere (Caimmi 
1980; 1983).   This discrepancy is probably 
owing to the fact, that (for an assigned 
generalized anisotropy
parameter) the isopycnic surfaces are 
(taken to be) rotationally distorted at 
the same extent for homeoidally striated
ellipsoids, while rotational distorsion
increases passing from the centre to the
boundary for self-gravitating fluids.

\subsection{Sequences of homeoidally
striated Jacobi ellipsoids}
\label{shsJe}

Sequences of homeoidally striated Jacobi
ellipsoids may be generated using 
Eqs.\,(\ref{seq:zief}) and (\ref{eq:Brszt}),
which make a system of three equations
involving five unknowns: the effective
anisotropy
parameters, $\tilde{\zeta}_{pp}$, and the
axis ratios, $\epsilon_{21}$ and
$\epsilon_{31}$.   If effective anisotropy
parameters are kept constant along the
sequence, then Eq.\,(\ref{eq:Brszt})
reduces to a relation involving axis
ratios only, and the rotation parameter,
$h$, may be determined using Eq.\,(\ref
{eq:upsh}).   If, on the other hand, the
rotation parameter is kept constant along
the sequence, then Eqs.\,(\ref{seq:zief}) 
and (\ref{eq:virh}) make a system of three   
equations in five unknowns, and the allowed
values of the effective anisotropy parameters,
or other rotation parameters, are represented
by a surface, $z=z(\epsilon_{21},\epsilon_{31})$.

In the special case of Jacobi ellipsoids,
incompressibility acts as an isotropic,
residual velocity distribution and ellipticities
are owing to rotation only, and the related
sequence is characterized by $\tilde{\zeta}_{pp}=
1/3$, $p=1,2,3$ (e.g., Jeans 1929, Chap.\,VIII;
Chandrasekhar 1966, Chaps.\,5-8).   It is true
also for gaseous, relaxed, homeoidally striated
Jacobi ellipsoids, where the collisional nature
of the fluid (in absence of systematic motions
different from rotation) necessarily yields an
isotropic residual velocity distribution, 
which coincides with Maxwell's distribution
(e.g., Jeans 1929, Chap.\,IX; James 1964).   On
the contrary, an anisotropic residual velocity
distribution may occur in presence of systematic
motions different from rotation within collisionless
fluids (e.g., Wiegandt 1982 a,b), such as stars or
non baryonic dark matter.

Generally speaking, there is no apparent criterion
for the selection of a specified sequence of
homeoidally striated Jacobi ellipsoids, instead of
one other.   To this respect, only the nature of
the special problem under consideration could
provide further information.    For this reason,
our attention shall be limited to sequences with
constant effective anisotropy parameters, $\tilde
{\zeta}_{pp}$, starting from a nonrotating,
axisymmetric configuration.   Accordingly, 
$\tilde{\zeta}_{11}=\tilde{\zeta}_{22}$, via
Eq.\,(\ref{eq:zp}).   The ending point is 
related to an oblong configuration, unless
centrifugal support at the ends of the
major equatorial axis occurs earlier.

\begin{figure}
\centering
\resizebox{\hsize}{!}{\includegraphics{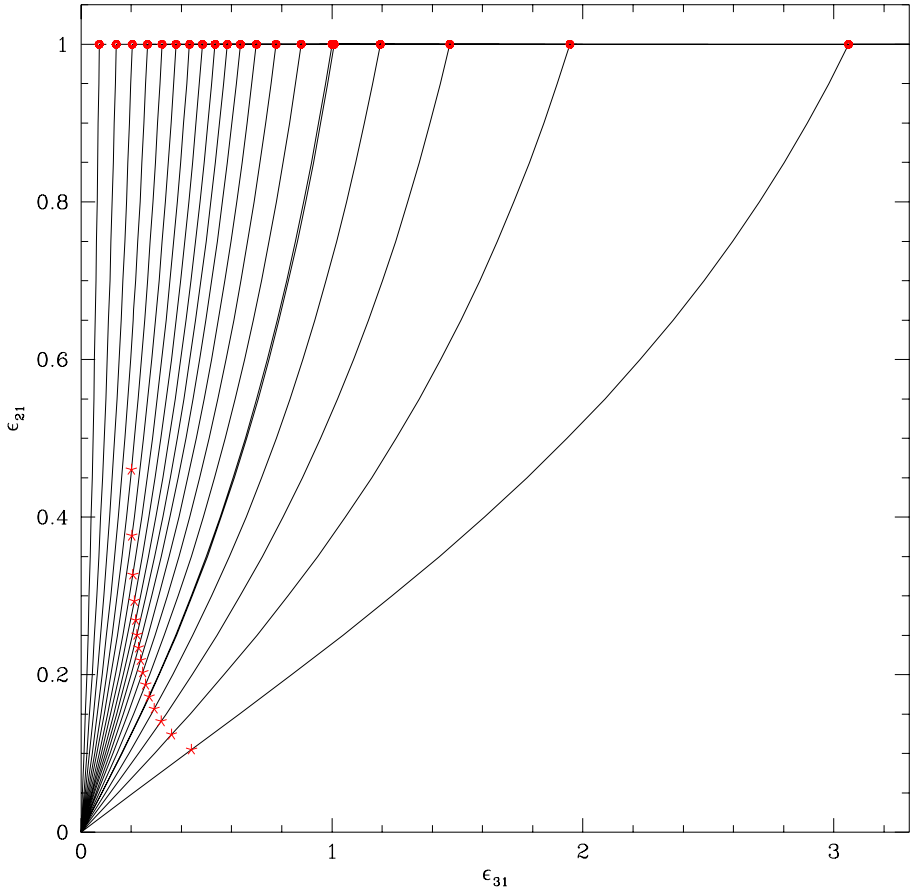}} 
\caption{The equatorial axis ratio, $\epsilon_ 
{21}$, as a function of the polar axis ratio,
$\epsilon_{31}$, for sequences of density
profiles in virial equilibrium starting from
nonrotating ($h=0$), spheroidal boundaries, 
with polar axis ratios $(\epsilon_{31})_{h=0}
=0.1~i$ (oblate) and $(\epsilon_{31})_{h=0}= 
1/(0.1~i)$ (prolate), $1\le i\le10$.   The
bifurcation from axisymmetric to triaxial
configurations is marked by a filled circle
on each curve.
The bifurcation from a round boundary takes
place in a sequence where $(\epsilon_{31})_
{h=0}=1.97698$.  A pentagon skeleton on each
curve marks the upper limit of polar
axis ratio, for which a necessary condition
for the occurrence of centrifugal support 
at the ends of the equatorial major axis, is
first satisfied with regard to a NFW density
profile (see Subsect.\,4.5) in rigid rotation.}
\label{f:epv}    
\end{figure}

The equatorial axis ratio, $\epsilon_{21}$,
as a function of the polar axis ratio,
$\epsilon_{31}$, for sequences of density
profiles of the kind considered, is plotted
in Fig.\,\ref{f:epv} with regard to initial,
nonrotating ($h=0$), spheroidal boundaries 
characterized 
by polar axis ratios within the range $0.1\le
(\epsilon_{31})_{h=0}\le10$.   
The bifurcation from axisymmetric to
triaxial configurations is marked by
a filled circle on each curve.   The
bifurcation from a round boundary takes
place in a sequence where $(\epsilon_
{31})_{h=0}=1.97698$.   Each sequence
is characterized by the initial,
nonrotating boundary, i.e. the effective
anisotropy parameters, $\tilde{\zeta}_{pp}$,
but it is independent of the special 
choice of density profile and (systematic
rotation) velocity profile.

The same holds for the rotation parameter, 
${\cal E}_{rot}$, defined by Eq.\,(\ref
{eq:Erz}).   With regard to the
remaining rotation parameters, let us
define their normalized counterparts,
via Eqs.\,(\ref{eq:hz}), (\ref{eq:upz}),
(\ref{eq:lam3}), and (\ref{eq:vrope}):
\begin{lefteqnarray}
\label{eq:hzN}
&& h_N=\frac{\nu_{ram}}{\nu_{sel}}h=\frac12
\frac{B_{sel}}{B_{ram}}\left[1-\frac{\zeta}
{\zeta_{33}}\frac{(B_{sel})_{33}}{B_{sel}}
\right]~~; \\
\label{eq:upzN}
&& \upsilon_N=\frac{\eta_{rot}\nu_{rot}}
{\nu_{sel}}\upsilon=\frac13\frac{\epsilon_
{21}\epsilon_{31}}{1+\epsilon_{21}^2}B_{sel}
\left[1-\frac{\zeta}{\zeta_{33}}\frac
{(B_{sel})_{33}}{B_{sel}}\right]~~; \\
\label{eq:lam3N}
&& \lambda_N^2=\frac{\nu_{ram}}{\nu_{sel}^2}
\frac{\eta_{rot}}{\eta_{anm}^2}\left[1-\frac
{\zeta^{-1}-1}{1+\zeta^{-1}\chi_v^2}\right]^
{-1}\lambda^2=\frac{1+\epsilon_{21}^2}4B_
{sel}^2\left[1-\frac\zeta{\zeta_{33}}\frac{(B_
{sel})_{33}}{B_{sel}}\right]~~; \\
\label{eq:chiN}
&& \chi_N^2=\frac{{\chi_v}^2}\zeta~~;
\end{lefteqnarray}
which are also independent of both 
density profile and (systematic rotation)
velocity profile.

\begin{figure}
   \centering
  \resizebox{\hsize}{!}{\includegraphics{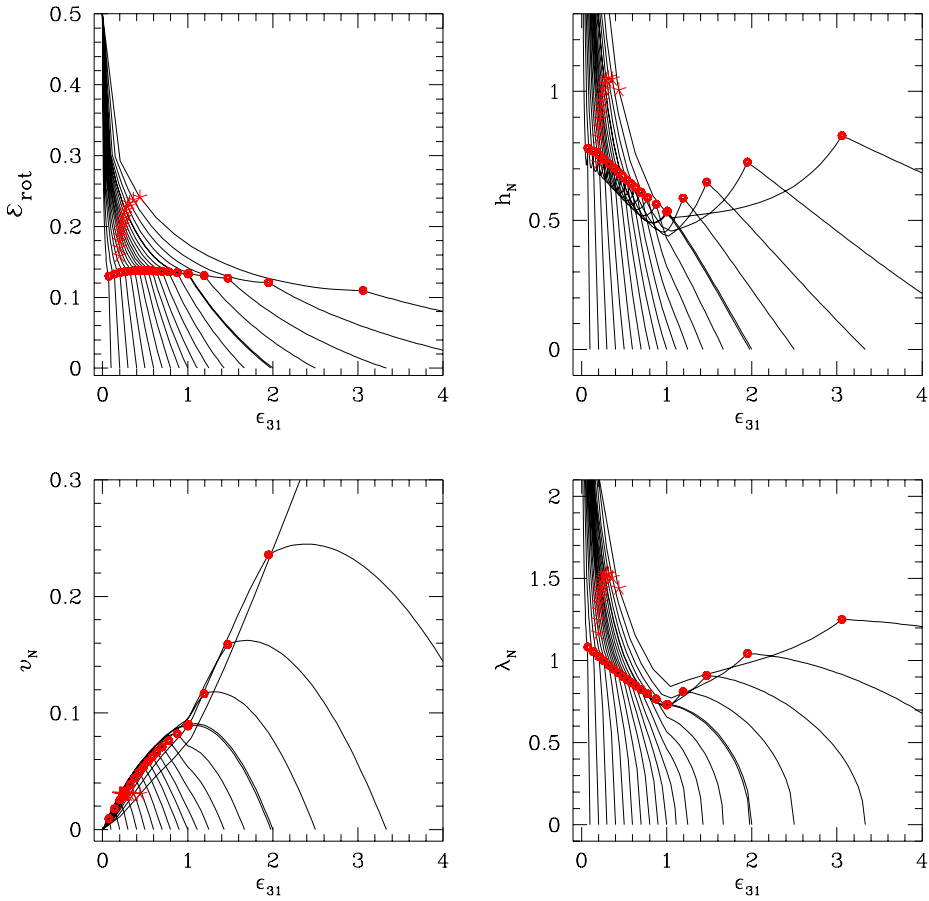}} 
\caption{The rotation parameters, ${\cal E}_{rot}$,
$h_N$, $\lambda_N$, and $\upsilon_N$ (from top 
left in clockwise sense), as a function of the
polar axis ratio, $\epsilon_{31}$, for
sequences of density profiles in virial 
equilibrium, starting from nonrotating   
($h=0$), spheroidal boundaries, 
with polar axis ratio $(\epsilon_{31})_{h=0}
=0.1~i$ (oblate) and $(\epsilon_{31})_{h=0}= 
1/(0.1~i)$ (prolate), $1\le i\le10$.  
Other captions as in Fig.\,\ref{f:epv}.
All the curves end at the oblong configuration,
where the axis ratios tend to zero.   The 
related values of the rotation parameters,  
${\cal E}_{rot}$ and $\upsilon_N$, tend to
0.5 and 0, respectively, while $h_N$ and
$\lambda_N$ tend to infinity.}
\label{f:pr3}    
\end{figure}

\begin{figure}
   \centering
  \resizebox{\hsize}{!}{\includegraphics{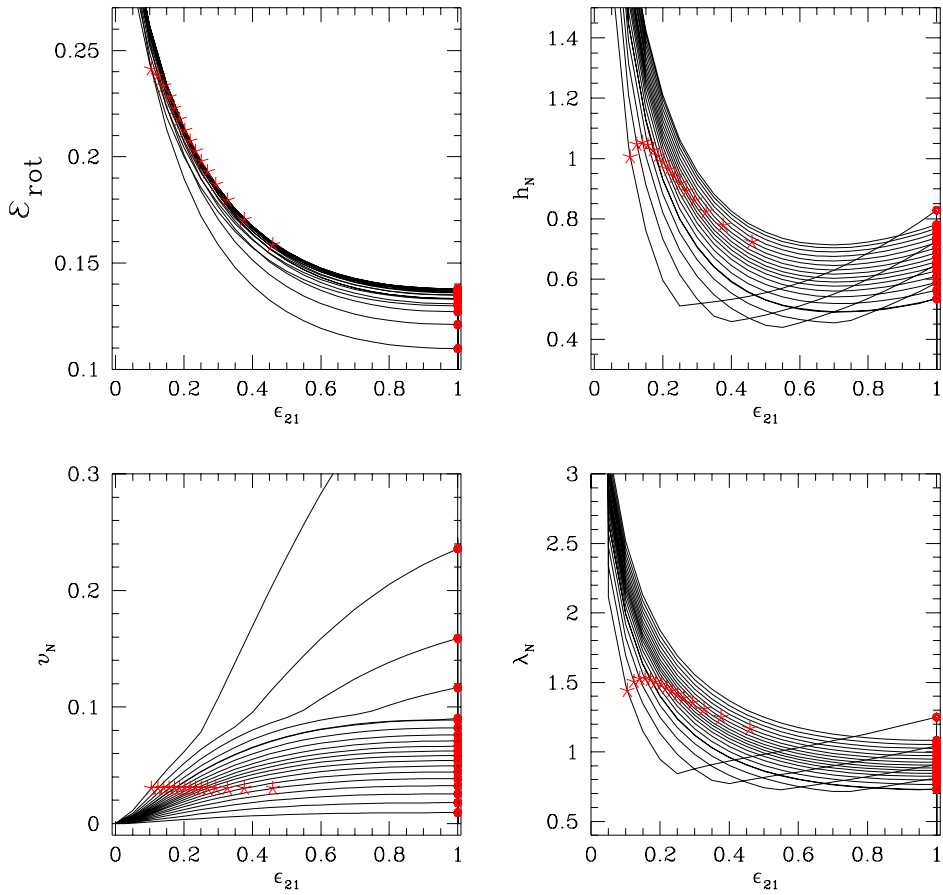}} 
\caption{The rotation parameters, ${\cal E}_{rot}$,
$h_N$, $\lambda_N$, and $\upsilon_N$ (from top 
left in clockwise sense), as a function of the
equatorial axis ratio, $\epsilon_{21}$, for
sequences of density profiles in virial 
equilibrium, starting from nonrotating   
($h=0$), spheroidal boundaries, 
with polar axis ratio $(\epsilon_{31})_{h=0}
=0.1~i$ (oblate) and $(\epsilon_{31})_{h=0}= 
1/(0.1~i)$ (prolate), $1\le i\le10$.  
Other captions as in Fig.\,\ref{f:epv}.
All the curves end at the oblong configuration,
where the axis ratios tend to zero.   The 
related values of the rotation parameters,  
${\cal E}_{rot}$ and $\upsilon_N$, tend to
0.5 and 0, respectively, while $h_N$ and
$\lambda_N$ tend to infinity.   The 
coordinates of the nonrotating configurations
are (1,0) in each panel.}
\label{f:pr2}    
\end{figure}

The rotation parameters, ${\cal E}_{rot}$,
$h_N$, $\lambda_N$, and $\upsilon_N$ (from top 
left in clockwise sense), as a function of 
the axis ratios, $\epsilon_{31}$ and $\epsilon
_{21}$, are plotted in Figs.\,\ref{f:pr3}
and \ref{f:pr2}, respectively, for sequences 
of density profiles in virial equilibrium 
starting from nonrotating ($h=0$), spheroidal 
boundaries, with polar axis ratios $(\epsilon_
{31})_{h=0}=0.1~i$ (oblate) and $(\epsilon_{31})
_{h=0}= 1/(0.1~i)$ (prolate), $1\le i\le10$.
Each curve is characterized by the effective
anisotropy parameters, $\tilde{\zeta}_{pp}$, but 
is independent of both density profile and
(systematic rotation) velocity profile.

\begin{figure}
\centering
\resizebox{\hsize}{!}{\includegraphics{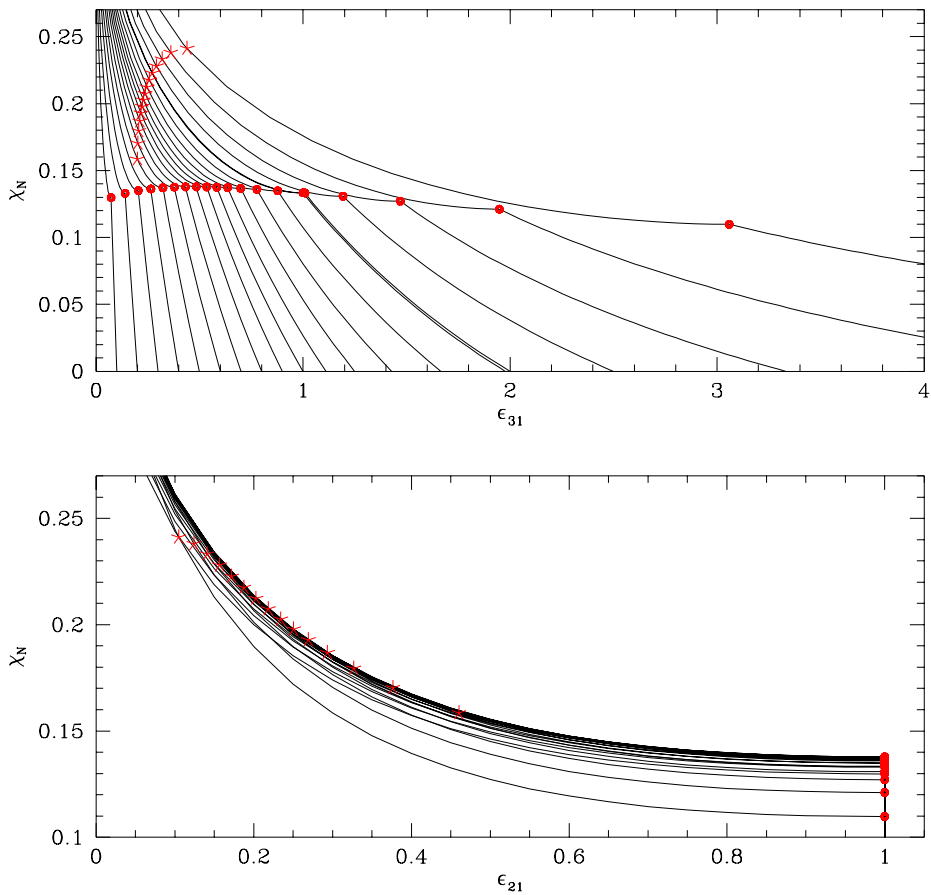}} 
\caption{The rotation parameter, $\chi_N$, as a 
function of the polar axis ratio, $\epsilon_{31}$ 
(upper panel) and the equatorial axis ratio, 
$\epsilon_ {21}$ (lower panel), for sequences of 
density profiles in virial equilibrium starting 
from nonrotating ($h=0$), spheroidal boundaries, 
with polar axis ratios $(\epsilon_{31})_{h=0}
=0.1~i$ (oblate) and $(\epsilon_{31})_{h=0}= 
1/(0.1~i)$ (prolate), $1\le i\le10$.   
Other captions as in Fig.\,\ref{f:epv}.
All the curves end at the oblong configuration,
where the axis ratios tend to zero and the 
rotation parameter, $\chi_N$, tends to  
infinity.   The coordinates of the starting
configuration related to all the sequences are
$(1,0)$ in the lower panel.}
\label{f:rvN}    
\end{figure}

The rotation parameter, $\chi_N$, as a function of 
the axis ratios, $\epsilon_{31}$ and $\epsilon
_{21}$, is plotted in Fig.\,\ref{f:rvN} for 
sequences of density profiles in virial equilibrium 
starting from nonrotating ($h=0$), spheroidal 
boundaries, with polar axis ratios $(\epsilon_
{31})_{h=0}=0.1~i$ (oblate) and $(\epsilon_{31})
_{h=0}= 1/(0.1~i)$ (prolate), $1\le i\le10$.

Axis ratios and normalized rotation parameters
related to virial equilibrium configurations
belonging to three typical sequences, are listed
in Tabs.\,\ref{t:sobi}, \ref{t:ssfi}, and
\ref{t:spri},
where the starting nonrotating configuration is
oblate (anisotropy induces flattening), spherical
(ispotropic case), and prolate (anisotropy induces
elongation), respectively.
\begin{table}
\begin{tabular}{llllllll}
\hline
\hline
\multicolumn{1}{c}{$\epsilon_{31}$} &
\multicolumn{1}{c}{$\epsilon_{21}$} &
\multicolumn{1}{c}{${\cal E}_{rot}$} &
\multicolumn{1}{c}{$h_N$} &
\multicolumn{1}{c}{$\upsilon_N$} &
\multicolumn{1}{c}{$\lambda_N$} &
\multicolumn{1}{c}{$\chi_N$} \\         
\hline
0.500000  & 1.00 & 0.0000000 & 0.000000 & 0.0000000 & 0.000000 & 0.000000 \\
0.450000  & 1.00 & 0.0355346 & 0.175722 & 0.0131792 & 0.466092 & 0.276598 \\
0.400000  & 1.00 & 0.0733964 & 0.371350 & 0.0247567 & 0.685355 & 0.414787 \\
0.350000  & 1.00 & 0.113834  & 0.589724 & 0.0344006 & 0.873944 & 0.542936 \\
{\bf0.322650} & {\bf1.00} & {\bf0.137144} & {\bf0.719984} & {\bf0.0387171} 
& {\bf0.972086} & {\bf0.614782} \\
0.314355  & 0.95 & 0.137244 & 0.703109 & 0.0386747 & 0.972967 & 0.615092 \\
0.305580  & 0.90 & 0.137567 & 0.688611 & 0.0385384 & 0.975802 & 0.616090 \\
0.296287  & 0.85 & 0.138149 & 0.676710 & 0.0382934 & 0.980919 & 0.617888 \\
0.286432  & 0.80 & 0.139034 & 0.667680 & 0.0379229 & 0.988715 & 0.620622 \\
0.275970  & 0.75 & 0.140272 & 0.661862 & 0.0374076 & 0.999672 & 0.624450 \\
0.264850  & 0.70 & 0.141925 & 0.659686 & 0.0367259 & 1.01438  & 0.629567 \\
0.253015  & 0.65 & 0.144066 & 0.661701 & 0.0358531 & 1.03357  & 0.636204 \\
0.240404  & 0.60 & 0.146785 & 0.668618 & 0.0347618 & 1.05816  & 0.644647 \\
0.226950  & 0.55 & 0.150191 & 0.681374 & 0.0334219 & 1.08933  & 0.655249 \\
0.212577  & 0.50 & 0.154418 & 0.701221 & 0.0318002 & 1.12859  & 0.668457 \\
0.197205  & 0.45 & 0.159634 & 0.729873 & 0.0298619 & 1.17795  & 0.684842 \\
0.180748  & 0.40 & 0.166054 & 0.769728 & 0.0275717 & 1.24013  & 0.705158 \\
0.163111  & 0.35 & 0.173953 & 0.824243 & 0.0248968 & 1.31896  & 0.730426 \\
0.144198  & 0.30 & 0.183700 & 0.898594 & 0.0218121 & 1.41997  & 0.762088 \\
0.123907  & 0.25 & 0.195803 & 1.00092  & 0.0183099 & 1.55171  & 0.802290 \\
0.102147  & 0.20 & 0.211000 & 1.14492  & 0.0144169 & 1.72824  & 0.854462 \\
0.0788405 & 0.15 & 0.230466 & 1.35619  & 0.0102269 & 1.97547  & 0.924692 \\
0.0539566 & 0.10 & 0.256360 & 1.69132  & 0.00596398 & 2.35031  & 1.02577 \\ 
0.0275656 & 0.05 & 0.294098 & 2.32663  & 0.00212718 & 3.02987  & 1.19513 \\
0.0000000 & 0.00 & 0.500000 &$+\infty$ & 0.00000000 &$+\infty$ & $+\infty$ \\
\hline\hline
\end{tabular}
\caption{Values of axis ratios, $\epsilon_{31}$ and
$\epsilon_{21}$, and normalized rotation parameters,
${\cal E}_{rot}$, $h_N$, $\upsilon_N$, $\lambda_N$,
$\chi_N$, related to a sequence of virial equilibrium
configurations with anisotropy parameters, $\tilde
{\zeta}_{11}=\tilde{\zeta}_{22}=0.391002$ and $\tilde
{\zeta}_{33}=0.217996$.   The polar axis ratio of the
initial nonrotating configuration is $\epsilon_{31}=
1/2$.   Values at the bifurcation point are listed
on the boldface line.}
\label{t:sobi}
\end{table}
\begin{table}
\begin{tabular}{llllllll}
\hline
\hline
\multicolumn{1}{c}{$\epsilon_{31}$} &
\multicolumn{1}{c}{$\epsilon_{21}$} &
\multicolumn{1}{c}{${\cal E}_{rot}$} &
\multicolumn{1}{c}{$h_N$} &
\multicolumn{1}{c}{$\upsilon_N$} &
\multicolumn{1}{c}{$\lambda_N$} &
\multicolumn{1}{c}{$\chi_N$} \\         
\hline
1.000000 & 1.00 & 0.0000000 & 0.0000000 & 0.00000000 & 0.000000 & 0.000000 \\
0.950000 & 1.00 & 0.0136423 & 0.0554973 & 0.00878707 & 0.237573 & 0.167481 \\
0.900000 & 1.00 & 0.0279335 & 0.115614  & 0.0173421  & 0.345874 & 0.243255 \\
0.850000 & 1.00 & 0.0429239 & 0.180831  & 0.0256177  & 0.436408 & 0.306447 \\
0.800000 & 1.00 & 0.0586689 & 0.251690  & 0.0335587  & 0.519556 & 0.364604 \\
0.750000 & 1.00 & 0.0752304 & 0.328809  & 0.0411011  & 0.599400 & 0.420843 \\
0.700000 & 1.00 & 0.0926773 & 0.412889  & 0.0481704  & 0.678136 & 0.476999 \\
0.650000 & 1.00 & 0.111086  & 0.504734  & 0.0546795  & 0.757185 & 0.534446  \\
0.600000 & 1.00 & 0.130544  & 0.605265  & 0.0605265  & 0.837600 & 0.594425  \\
{\bf0.582724} & {\bf1.00} & {\bf0.137528}  & {\bf0.642207}  & {\bf0.0623716}  
& {\bf0.865863} & {\bf0.615969}  \\
0.567738 & 0.95 & 0.137631  & 0.627187  & 0.0623058  & 0.866686 & 0.616287  \\
0.551873 & 0.90 & 0.137962  & 0.614355  & 0.0620944  & 0.869331 & 0.617310  \\
0.535058 & 0.85 & 0.138559  & 0.603917  & 0.0617145  & 0.874109 & 0.619154  \\
0.517216 & 0.80 & 0.139466  & 0.596127  & 0.0611395  & 0.881390 & 0.621957  \\
0.498264 & 0.75 & 0.140736  & 0.591305  & 0.0603394  & 0.891630 & 0.625885  \\
0.478110 & 0.70 & 0.142431  & 0.589855  & 0.0592799  & 0.905387 & 0.631136  \\
0.456652 & 0.65 & 0.144629  & 0.592298  & 0.0579219  & 0.923359 & 0.637951  \\
0.433781 & 0.60 & 0.147422  & 0.599309  & 0.0562217  & 0.946423 & 0.646627  \\
0.409377 & 0.55 & 0.150922  & 0.611785  & 0.0541300  & 0.975705 & 0.657530  \\
0.383309 & 0.50 & 0.155270  & 0.630931  & 0.0515928  & 1.01267  & 0.671126  \\
0.355436 & 0.45 & 0.160640  & 0.658405  & 0.0485518  & 1.05927  & 0.688014  \\
0.325609 & 0.40 & 0.167257  & 0.696539  & 0.0449463  & 1.11817  & 0.708986  \\
0.293667 & 0.35 & 0.175409  & 0.748717  & 0.0407171  & 1.19312  & 0.735119  \\
0.259447 & 0.30 & 0.185482  & 0.820038  & 0.0358145  & 1.28960  & 0.767941  \\
0.222783 & 0.25 & 0.198006  & 0.918565  & 0.0302122  & 1.41609  & 0.809731  \\
0.183524 & 0.20 & 0.213754  & 0.105795  & 0.0239349  & 1.58664  & 0.864147  \\
0.141550 & 0.15 & 0.233939  & 0.126381  & 0.0171106  & 1.82719  & 0.937692  \\
0.968152 & 0.10 & 0.260769  & 0.159292  & 0.0100787  & 2.19478  & 1.04405   \\
0.494435 & 0.05 & 0.299688  & 0.222228  & 0.00364434 & 2.86687  & 1.22315   \\
0.000000 & 0.00 & 0.500000  & $+\infty$ & 0.00000000 & $+\infty$ & $+\infty$ \\
\hline\hline                                                                
\end{tabular}                                                               
\caption{Values of axis ratios, $\epsilon_{31}$ and                         
$\epsilon_{21}$, and normalized rotation parameters,
${\cal E}_{rot}$, $h_N$, $\upsilon_N$, $\lambda_N$,
$\chi_N$, related to a sequence of virial equilibrium
configurations with anisotropy parameters, $\tilde
{\zeta}_{11}=\tilde{\zeta}_{22}=\tilde{\zeta}_{33}=1/3$.   
The polar axis ratio of the
initial nonrotating configuration is $\epsilon_{31}=
1$.   Values at the bifurcation point are listed
on the boldface line.}
\label{t:ssfi}
\end{table}
\begin{table}
\begin{tabular}{llllllll}
\hline
\hline
\multicolumn{1}{c}{$\epsilon_{31}$} &
\multicolumn{1}{c}{$\epsilon_{21}$} &
\multicolumn{1}{c}{${\cal E}_{rot}$} &
\multicolumn{1}{c}{$h_N$} &
\multicolumn{1}{c}{$\upsilon_N$} &
\multicolumn{1}{c}{$\lambda_N$} &
\multicolumn{1}{c}{$\chi_N$} \\         
\hline
2.00000   & 1.00 & 0.00000000 & 0.0000000 & 0.00000000 & 0.000000 & 0.0000000 \\
1.95000   & 1.00 & 0.00478379 & 0.0286995 & 0.00932733 & 0.207471 & 0.0982853 \\
1.90000   & 1.00 & 0.00971323 & 0.0574460 & 0.0181912  & 0.291439 & 0.140753  \\
1.85000   & 1.00 & 0.0147955  & 0.0862190 & 0.0265842  & 0.354412 & 0.174624  \\
1.80000   & 1.00 & 0.0200384  & 0.114995  & 0.0344986  & 0.406180 & 0.204328  \\
1.75000   & 1.00 & 0.0254502  & 0.143749  & 0.0419268  & 0.450536 & 0.231582  \\
1.70000   & 1.00 & 0.0310397  & 0.172451  & 0.0488610  & 0.489413 & 0.257271  \\
1.65000   & 1.00 & 0.0368164  & 0.201067  & 0.0552935  & 0.523951 & 0.281932  \\
1.60000   & 1.00 & 0.0427906  & 0.229562  & 0.0612164  & 0.554875 & 0.305926  \\
1.55000   & 1.00 & 0.0489731  & 0.257892  & 0.0666221  & 0.582678 & 0.329517  \\
1.50000   & 1.00 & 0.0553759  & 0.286012  & 0.0715029  & 0.607705 & 0.352910  \\
1.45000   & 1.00 & 0.0620118  & 0.313867  & 0.0758512  & 0.630201 & 0.376275  \\
1.40000   & 1.00 & 0.0688944  & 0.341399  & 0.0796598  & 0.650340 & 0.399761  \\
1.35000   & 1.00 & 0.0760387  & 0.368540  & 0.0829216  & 0.668248 & 0.423501  \\
1.30000   & 1.00 & 0.0834610  & 0.395215  & 0.0856298  & 0.684008 & 0.447625  \\
1.25000   & 1.00 & 0.0911788  & 0.421336  & 0.0877783  & 0.697672 & 0.472259  \\
1.20000   & 1.00 & 0.0992113  & 0.446806  & 0.0893613  & 0.709265 & 0.497534  \\
1.15000   & 1.00 & 0.107579   & 0.471516  & 0.0903739  & 0.718791 & 0.523586   \\
1.10000   & 1.00 & 0.116305   & 0.495340  & 0.0908123  & 0.726229 & 0.550563   \\
1.05000   & 1.00 & 0.125415   & 0.518135  & 0.0906737  & 0.731541 & 0.578628   \\
{\bf1.00884}   & {\bf1.00} & {\bf0.133221}   & {\bf0.536015}  & {\bf0.0901254}  
& {\bf0.734279} & {\bf0.602676}   \\
0.982862  & 0.95 & 0.133325   & 0.518938  & 0.0892464  & 0.728588 & 0.602997   \\
0.955297  & 0.90 & 0.133660   & 0.508463  & 0.0889593  & 0.730978 & 0.604029   \\
0.926018  & 0.85 & 0.134263   & 0.500075  & 0.0884430  & 0.735296 & 0.605890   \\
0.894889  & 0.80 & 0.135180   & 0.494001  & 0.0876613  & 0.741881 & 0.608720   \\
0.861761  & 0.75 & 0.136465   & 0.490528  & 0.0865726  & 0.751151 & 0.612687   \\
0.826476  & 0.70 & 0.138183   & 0.490021  & 0.0851292  & 0.763622 & 0.617993   \\
0.788860  & 0.65 & 0.140412   & 0.492953  & 0.0832766  & 0.779942 & 0.624884   \\
0.748729  & 0.60 & 0.143247   & 0.499946  & 0.0809524  & 0.800933 & 0.633663   \\
0.705885  & 0.55 & 0.146806   & 0.511829  & 0.0780864  & 0.827656 & 0.644711   \\
0.660119  & 0.50 & 0.151235   & 0.529730  & 0.0745994  & 0.861506 & 0.658506   \\
0.611212  & 0.45 & 0.156719   & 0.555209  & 0.0704044  & 0.904353 & 0.675673   \\
0.558939  & 0.40 & 0.163494   & 0.590488  & 0.0654076  & 0.958773 & 0.697036   \\
0.503073  & 0.35 & 0.171870   & 0.638821  & 0.0595133  & 1.02842  & 0.723729   \\
0.443394  & 0.30 & 0.182258   & 0.705156  & 0.0526322  & 1.11870  & 0.757365   \\
0.379704  & 0.25 & 0.195231   & 0.797402  & 0.0447006  & 1.23801  & 0.800368   \\
0.311845  & 0.20 & 0.211623   & 0.929083  & 0.0357161  & 1.40036  & 0.856647   \\
0.239733  & 0.15 & 0.232741   & 1.12576   & 0.0258135  & 1.63179  & 0.933190   \\
0.163419  & 0.10 & 0.260936   & 1.44447   & 0.0154269  & 1.98960  & 1.04474    \\
0.0832065 & 0.05 & 0.301873   & 2.06324   & 0.00569398 & 2.65204  & 1.23435    \\
0.0000000 & 0.00 & 0.500000   &$+\infty$  & 0.00000000 &$+\infty$ & $+\infty$   \\
\hline\hline                                                                
\end{tabular}                                                               
\caption{Values of axis ratios, $\epsilon_{31}$ and                         
$\epsilon_{21}$, and normalized rotation parameters,
${\cal E}_{rot}$, $h_N$, $\upsilon_N$, $\lambda_N$,
$\chi_N$, related to a sequence of virial equilibrium
configurations with anisotropy parameters, $\tilde
{\zeta}_{11}=\tilde{\zeta}_{22}=0.27173$, $\tilde
{\zeta}_{33}=
0.45654$.   The polar axis ratio of the
initial nonrotating configuration is $\epsilon_{31}=
2$.   Values at the bifurcation point are listed
on the boldface line.}
\label{t:ssfi}
\end{table}
\begin{table}
\begin{tabular}{lllllllll}
\hline
\hline
\multicolumn{1}{c}{$\tilde{\zeta}_{33}$} &
\multicolumn{1}{c}{$(\epsilon_{31})_i$} &
\multicolumn{1}{c}{$(\epsilon_{31})_b$} &
\multicolumn{1}{c}{$({\cal E}_{rot})_b$} &
\multicolumn{1}{c}{$(h_N)_b$} &
\multicolumn{1}{c}{$(\upsilon_N)_b$} &
\multicolumn{1}{c}{$(\lambda_N)_b$} &
\multicolumn{1}{c}{$(\chi_N)_b$} \\         
\hline
0.0582397 & 0.10000 & 0.722934 & 0.129818 & 0.780139 & 0.00939982 & 1.08262  & 0.592188 \\ 
0.107390  & 0.20000 & 0.139949 & 0.132948 & 0.768228 & 0.0179188  & 1.05346  & 0.601834 \\ 
0.149487  & 0.30000 & 0.203842 & 0.134996 & 0.753170 & 0.0255879  & 1.02495  & 0.608151 \\ 
0.185995  & 0.40000 & 0.264590 & 0.136320 & 0.736767 & 0.0324902  & 0.997747 & 0.612238 \\ 
0.217996  & 0.50000 & 0.322650 & 0.137144 & 0.719984 & 0.0387171  & 0.972086 & 0.614782 \\ 
0.246304  & 0.60000 & 0.378373 & 0.137614 & 0.703346 & 0.0443546  & 0.948000 & 0.616233 \\ 
0.271548  & 0.70000 & 0.432038 & 0.137828 & 0.687137 & 0.0494782  & 0.925433 & 0.616894 \\ 
0.294221  & 0.80000 & 0.483866 & 0.137855 & 0.671506 & 0.0541532  & 0.904293 & 0.616977 \\ 
0.314711  & 0.90000 & 0.534044 & 0.137743 & 0.656521 & 0.0584352  & 0.884471 & 0.616632 \\ 
0.333333  & 1.00000 & 0.582724 & 0.137528 & 0.642207 & 0.0623716  & 0.865863 & 0.615969 \\ 
0.342473  & 1.05263 & 0.607789 & 0.137383 & 0.634943 & 0.0643185  & 0.856522 & 0.615519 \\ 
0.352143  & 1.11111 & 0.635215 & 0.137200 & 0.627086 & 0.0663890  & 0.846487 & 0.614954 \\ 
0.362396  & 1.17647 & 0.665365 & 0.136973 & 0.618566 & 0.0685954  & 0.835677 & 0.614254 \\ 
0.373286  & 1.25000 & 0.698684 & 0.136694 & 0.609302 & 0.0709515  & 0.824000 & 0.613394 \\ 
0.384880  & 1.33333 & 0.735717 & 0.136355 & 0.599199 & 0.0734735  & 0.811345 & 0.612345 \\ 
0.397253  & 1.42857 & 0.777153 & 0.135943 & 0.588145 & 0.0761798  & 0.797584 & 0.611072 \\ 
0.410492  & 1.53846 & 0.823862 & 0.135443 & 0.576007 & 0.0790917  & 0.782562 & 0.609532 \\ 
0.424701  & 1.66667 & 0.876970 & 0.134839 & 0.562626 & 0.0822344  & 0.766093 & 0.607668 \\
0.440001  & 1.81818 & 0.937962 & 0.134108 & 0.547809 & 0.0856374  & 0.747948 & 0.605413 \\
0.454545  & 1.97698 & 1.00000  & 0.133333 & 0.533331 & 0.0888886  & 0.730295 & 0.603021 \\
0.456540  & 2.00000 & 1.00884  & 0.133221 & 0.536015 & 0.0901254  & 0.734279 & 0.602676 \\   
0.474496  & 2.22222 & 1.09237  & 0.132139 & 0.560233 & 0.101997   & 0.770589 & 0.599341 \\   
0.494095  & 2.50000 & 1.19251  & 0.130812 & 0.586789 & 0.116626   & 0.811199 & 0.595251 \\
0.515624  & 2.85714 & 1.31517  & 0.129170 & 0.616069 & 0.135040   & 0.857074 & 0.590193 \\
0.539464  & 3.33333 & 1.46959  & 0.127114 & 0.648565 & 0.158854   & 0.909552 & 0.583859 \\
0.566145  & 4.00000 & 1.67120  & 0.124494 & 0.684923 & 0.190774   & 0.970594 & 0.575792 \\
0.596454  & 5.00000 & 1.94831  & 0.121076 & 0.726026 & 0.235754   & 1.04326  & 0.565265 \\
0.631682  & 6.66667 & 2.35994  & 0.116446 & 0.773166 & 0.304104   & 1.13287  & 0.550997 \\
0.674330  &10.0000  & 3.05847  & 0.109765 & 0.828464 & 0.422306   & 1.25029  & 0.530357 \\
0.731036  &20.0000  & 4.64473  & 0.0986983& 0.896310 & 0.693852   & 1.42650  & 0.495929 \\
\hline\hline
\end{tabular}                                                                           
\caption{Values of effective anisotropy parameters,
$\tilde{\zeta}_{33}$, polar axis ratios of non rotating 
configurations, $(\epsilon_{31})_i$, and polar axis ratios,                                     
$(\epsilon_{31})_b$, and normalized rotation parameters,                                    
$({\cal E}_{rot})_b$, $(h_N)_b$, $(\upsilon_N)_b$, 
$(\lambda_N)_b$, $(\chi_N)_b$, of configurations at
the bifurcation point, related to sequences of virial 
equilibrium configurations with constant effective 
anisotropy parameters.   Values at the isotropic 
($\tilde{\zeta}_{11}=\tilde{\zeta}_{22}=\tilde{\zeta}_{33}$)
are listed on the boldface line.}
\label{t:seib}
\end{table}

\subsection{A necessary condition for centrifugal
support}\label{ncfcs}

With regard to Jacobi ellipsoids, rotational
equilibrium i.e. balance between gravitational
and centrifugal force, first occurs at the end
of the major equatorial semiaxis, for flat or oblong
configurations (e.g., Caimmi 1980).   On the
other hand, homeoidally striated
Jacobi ellipsoids show rotational equilibrium
for less flattened or less elongated configurations. 
The calculation of
the gravitational force induced by density
profiles of the kind considered, involves
numerical integrations (e.g., Chandrasekhar
1969, Chap.\,3, \S 20) and is outside the aim
of the current attempt.   

Our attention shall 
be restricted here to two limiting situations,
for which the calculation of the gravitational
force induced by the mass distribution is simple,
and the corresponding range includes the
gravitational force induced by the system
under investigation.   In
addition, our considerations will be restricted
to the boundary, which is the case of physical
interest.   On the other hand, the procedure
may be extended to a generic, isopycnic surface.

Given a homeoidally striated, Jacobi ellipsoid,
let us define a related, striated sphere and
focaloidally striated ellipsoid, as matter
distributions with same scaled density profile,
total mass, 
major equatorial semiaxis and, in the latter case, same
boundary.   Owing to MacLaurin's theorem
(e.g., Chandrasekhar 1969, Chap.\,3, \S 19),
the gravitational force induced by the
striated sphere and the focaloidally striated
ellipsoid at the end of the major equatorial semiaxis, is:
\begin{equation}
\label{eq:FG}
F_G(a_1,0,0)=-2\pi G\bar{\rho}_w(A_1)_wa_1~~;
\end{equation}
where $\bar{\rho}$ is the mean density, $A_1$
is a shape factor in the range $0\le A_1\le2/3$
with the upper limit related to spherical
configurations, and $w=sph,\,foc,$ for the
striated sphere and the focaloidally striated
ellipsoid, respectively.

The balance between gravitational and centrifugal
force at the end of the major equatorial semiaxis, reads:
\begin{equation}
\label{eq:eqr}
-2\pi G\bar{\rho}_w(A_1)_wa_1+\Omega^2a_1=0~~;
\end{equation}
which, using Eq.\,(\ref{eq:vg}), may be cast
under the equivalent form:
\begin{equation}
\label{eq:eqv}
(\upsilon_{eq})_w=\frac{\Omega^2}{2\pi G\bar{\rho}
_w}=(A_1)_w~~;
\end{equation}
where the index, $eq$, denotes the
rotation parameter, $\upsilon$, related to
rotational equilibrium at the end of major
equatorial semiaxis.   The above results allow the
validity of the relation:
\begin{equation}
\label{eq:inev}
(\upsilon_{eq})_{foc}\le\upsilon_{eq}\le
(\upsilon_{eq})_{sph}~~;
\end{equation}
where $\upsilon_{eq}$ is the critical 
value related to the homeoidally striated,
Jacobi ellipsoid.   Then $\upsilon\ge
(\upsilon_{eq})_{foc}$ and $\upsilon\le
(\upsilon_{eq})_{sph}$ make a sufficient
and a necessary condition, respectively,
for the occurrence of rotational equilibrium
at the end of major equatorial semiaxis, in 
homeoidally striated, Jacobi ellipsoids.   

Owing to MacLaurin's theorem, the condition of
rotational equilibrium, Eq.\,(\ref{eq:eqv}),
for focaloidally striated ellipsoids coincides
with its counterpart related to homogeneous
ellipsoids with equal mean density, axis ratios,
and velocity field.   On the other hand, the
rotation parameter, $\upsilon$, in the special 
case of focaloidally striated ellipsoids, is
also expressed by Eq.\,(\ref{eq:upsh}), 
particularized to Jacobi ellipsoids, owing to
MacLaurin's theorem.   A
comparison with Eq.\,(\ref{eq:eqv})
shows that rotational equilibrium in
focaloidally striated ellipsoids occurs
only for flat configurations, $\epsilon_
{31}=0$.   Accordingly,
Eq.\,(\ref{eq:inev}) reduces to:
\begin{equation}
\label{eq:eqe31}
0\le\upsilon_{eq}\le\frac23~~;
\end{equation}
owing to Eq.\,(\ref{eq:eqv}) and $(A_1)_
{sph}=2/3$, as reported in Appendix B.

\section{Transitions from and towards
homeoidally striated density profiles}
\label{trans}

As an application of the results of Sect.\,3,
let us take into consideration the transition
from an initial configuration, with assigned
mass, angular momentum, and different kinds
of energy, towards a final configuration with 
assigned mass, angular momentum, and different
kinds of energy tensors.   Let both initial
and final configurations be homeoidally 
striated ellipsoids, with assigned density
profile and rotational velocity profile of
the kind discussed in Sect.\,\ref{gente}.
The above mentioned assumption is only
for sake of simplicity.   It is worth
emphasizing that the procedure outlined in
this Section, well holds for any density
profile with triplanar symmetry, i.e. 
symmetric with respect to the principal
planes.

The original method conceived by Thuan \&
Gott (1975) is aimed to be generalized on
many respects, namely (i) from spheroidal 
to ellipsoidal configurations; (ii) from 
isotropic to anisotropic, residual velocity
distributions; (iii) from relaxed to 
unrelaxed, final configurations; and, in
the limit of axisymmetric configurations
(iv) from rigid to differential rotation.

In the following, first the
general case shall be dealt with, in
connection with the transitions of the
kind considered, and then a few interesting
special cases will be analysed.

\subsection{The general case}
\label{trang}

The investigation of unrelaxed configurations,
for which virial equilibrium does not coincide
with dynamical (or hydrostatic) equilibrium, goes
beyond a purely academic interest.
For instance, (non interacting) galaxies
and cluster of galaxies (where galaxies
are still falling in from outer regions)
make examples of (approximately) relaxed and unrelaxed
configurations, respectively.

In general, the tensor components,
$\tilde{\zeta}_{pp}=\zeta_{pp}/\zeta$ and 
$(\tilde
{E}_{res})_{pq}$, which appear in Eq.\,(\ref
{seq:zirf}), have to be considered as 
effective, instead of actual, anisotropy 
parameters and 
residual kinetic-energy tensor, 
respectively.   Then the coincidence of
virial equilibrium with dynamical (or
hydrostatic) equilibrium
is restricted to the special choice $\zeta=1$
which, on the other hand, makes a necessary
condition only.
It is worth recalling that the residual 
kinetic energy, $E_
{res}$, has to be intended as related to
all motions different from systematic
rotation around a fixed axis, such as
random motions, streaming flows, coherent
oscillations, and so on.   
With regard to the initial configuration,
the residual kinetic energy will be expressed
as the sum of two contributions: one, owing
to systematic motions (mainly radial expansion),
and one other, owing to random motions.

For sake of brevity, let us define the
tensors, together with the related scalar 
quantities:
\begin{lefteqnarray}
\label{eq:Spq}
&& {\cal S}_{pq}=\nu_{sel}(B_{sel})_{pq}~~;
\quad{\cal S}=\nu_{sel}B_{sel}~~; \\
\label{eq:Rpq}
&& {\cal R}_{pq}=\nu_{ram}(B_{ram})_{pq}~~;
\quad{\cal R}=\nu_{ram}B_{ram}~~;
\end{lefteqnarray}
and let a prime, and its absence, denote 
the initial and the final configuration, 
respectively.   The effective
anisotropy parameters, $\tilde{\zeta}_{pp}$, 
expressed by Eqs.\,(\ref{eq:zq}),
and the rotation parameters, $h$, 
${\cal E}_{rot}$, $\upsilon$, $\lambda$, 
and $\chi_v$, expressed by 
Eqs.\,(\ref{eq:hz}), (\ref{eq:Erz}),
(\ref{eq:upz}), (\ref{eq:lam3}), and
(\ref{eq:vrope}), respectively, are 
written in compact notation in Appendix 
C, using Eqs.\,(\ref{eq:Spq}) 
and (\ref{eq:Rpq}).

Owing to 
Eqs.\,(\ref{eq:Espq}), (\ref{seq:ErJ}), 
(\ref{eq:Spq}), and (\ref{eq:Rpq}), the
self-potential and rotational energy
related to the final and initial
configuration, respectively, are:
\begin{lefteqnarray}
\label{eq:tEse}
&& E_{sel}=-\frac{GM^2}{a_1}{\cal S}~~;
\quad E_{sel}^\prime=-\frac{GM^{\prime2}}
{a_1^\prime}{\cal S}^\prime~~; \\
\label{eq:tEro}
&& E_{rot}=\frac{J^2}{Ma_1^2}{\cal R}~~;
\quad E_{rot}^\prime=\frac{J^{\prime2}}
{M^\prime a_1^{\prime2}}{\cal R}^\prime~~;
\end{lefteqnarray}
and the identity:
\begin{displaymath}
E_{rot}=\frac{E_{rot}}{E_{rot}^\prime}
\frac{E_{rot}^\prime}{-E_{sel}^\prime}
\frac{E_{sel}^\prime}{E_{sel}}(-E_{sel})~~;
\end{displaymath}
owing to Eqs.\,(\ref{seq:ErE}), (\ref
{eq:tEse}), and (\ref{eq:tEro}), may be
cast under the equivalent form:
\begin{leftsubeqnarray}
\slabel{eq:teada}
&& \frac{a_1^\prime}{a_1}=\frac
{\beta_M^3}{\beta_J^2}\frac{\cal S}
{{\cal S}^\prime}\frac{{\cal R}^\prime}
{\cal R}\frac{{\cal E}_{rot}}{{\cal E}_
{rot}^\prime}~~; \\
\slabel{eq:teadb}
&& \beta_J=\frac J{J^\prime}~~;\quad\beta_M=
\frac M{M^\prime}~~;
\label{seq:tead}
\end{leftsubeqnarray}
involving dimensionless quantities only.

On the other hand, the identity:
\begin{displaymath}
E=E_{sel}+E_{rot}+E_{res}=\frac E{E^
\prime}E^\prime=\frac E{E^\prime}(E_{sel}^
\prime+E_{rot}^\prime+E_{pec}^\prime+E_{osc}
^\prime)~~;
\end{displaymath}
by use of Eqs.\,(\ref{eq:virte}), and
(\ref{seq:zirf}), may be cast under the form:
\begin{leftsubeqnarray}
\slabel{eq:tenea}
&& \frac{E_{sel}}{E_{sel}^\prime}-{\cal E}_{rot}
\frac{E_{sel}}{E_{sel}^\prime}-\frac1{2\zeta}
\frac{E_{sel}}{E_{sel}^\prime}+\frac1\zeta\frac
{E_{sel}}{E_{sel}^\prime}{\cal E}_{rot}=\beta_E
(1-{\cal E}_{rot}^\prime-{\cal E}_{pec}^\prime-
{\cal E}_{osc}^\prime)~~; \\
\slabel{eq:teneb}
&& \beta_E=\frac E{E^\prime}~~;\quad{\cal E}
_{rot}=-\frac{E_{rot}}{E_{sel}}~~;\quad{\cal 
E}_{pec}=-\frac{E_{pec}}{E_{sel}}~~;\quad
{\cal E}_{osc}=-\frac{E_{osc}}{E_{sel}}~~; \\
\slabel{eq:tenec}
&& \zeta=-\frac{E_{sel}+2E_{rot}}{2(E_{osc}
+E_{pec})}~~;
\label{seq:tene}
\end{leftsubeqnarray}
where $E_{osc}$, $E_{pec}$, represent the 
kinetic energy of systematic radial and 
non systematic motions, respectively.
The combination of Eqs.\,(\ref{eq:tEse}), 
(\ref{eq:tEro}), and (\ref{seq:tene}) 
yields after some algebra:
\begin{lefteqnarray}
\label{eq:tead}
&& \beta_M^2\frac{{\cal S}}{{\cal S}^
\prime}\frac{a_1^\prime}{a_1}\left[\frac
{2\zeta-1}{2\zeta}-\frac{\zeta-1}\zeta
{\cal E}_{rot}\right]-\beta_E(1-{\cal E}
_{rot}^\prime-{\cal E}_{pec}^\prime-
{\cal E}_{osc}^\prime)=0~~;
\end{lefteqnarray}
involving dimensionless quantities only.

The combination of Eqs.\,(\ref
{eq:rErs}), (\ref{eq:Spq}), (\ref{eq:Rpq}), 
(\ref{eq:tEse}), and (\ref{eq:tEro})
yields after some algebra:
\begin{lefteqnarray}
\label{eq:tra}
&& {\cal E}_{rot}=\frac{{\cal R}}{{\cal S}}h
=\frac12\frac{{\cal R}}{{\cal S}}\frac{{\cal 
S}_{qq}}{{\cal R}_{qq}}\left[1-\frac{\zeta_{qq}}
{\zeta_{33}}\frac{{\cal S}_{33}}{{\cal S}_{qq}}
\right]~~;\quad q=1,2~~;
\end{lefteqnarray}
which, owing to Eq.\,(\ref{eq:Erz}), is
equivalent to:
\begin{lefteqnarray}
\label{eq:tro}
&& {\cal E}_{rot}=\frac12\left[1-\frac\zeta
{\zeta_{33}}\frac{{\cal S}_{33}}{{\cal S}}
\right]~~;
\end{lefteqnarray}
and Eq.\,(\ref{eq:Brszt}) may be written
under the equivalent form:
\begin{equation}
\label{eq:RSz}
\zeta_{33}\left[{\cal S}_{11}{\cal R}_{22}-
{\cal S}_{22}{\cal R}_{11}\right]=
{\cal S}_{33}\left[\zeta_{11}{\cal R}_{22}-
\zeta_{22}{\cal R}_{11}\right]~~;
\end{equation}
where Eqs.\,(\ref{eq:Spq}) and (\ref
{eq:Rpq}) have also been used.   The
substitution of Eq.\,(\ref{eq:teada}) 
into (\ref{eq:tead}) produces a second-degree 
equation in ${\cal E}_{rot}^\prime$, as:
\begin{leftsubeqnarray}
\slabel{eq:tre2a}
&& {\cal E}_{rot}^{\prime2}-2b{\cal E}_{rot}
^\prime+c=0~~; \\
\slabel{eq:tre2b}
&& b=\frac12\left(1-
{\cal E}_{osc}^\prime-
{\cal E}_{pec}^\prime\right)~~; \\
\slabel{eq:tre2c}
&& c=\frac{\beta_M^5}{\beta_J^2\beta_E}
\left(\frac{{\cal S}}{{\cal S}^\prime}
\right)^2\frac{{{\cal R}^\prime}}{\cal R}
{\cal E}_{rot}\left[\frac{2\zeta-1}{2\zeta}
-\frac{\zeta-1}\zeta{\cal E}_{rot}\right]
~~;
\label{seq:tre2}
\end{leftsubeqnarray}
the (reduced) discriminant of this equation is:
\begin{lefteqnarray}
\label{eq:del}
&& \Delta=b^2-\frac{\beta_M^5}{\beta_J^2\beta_E}
\left(\frac{{\cal S}}{{\cal S}^\prime}\right)^2
\frac{{\cal R}^\prime}{{\cal R}}{\cal E}_
{rot}\left[\frac{2\zeta-1}{2\zeta}
-\frac{\zeta-1}\zeta{\cal E}_{rot}\right]
~~;
\end{lefteqnarray}
with regard to a transition from an initial
to a final configuration, where all the
parameters which appear in Eq.\,(\ref{eq:del})
are specified, except the axis ratios, $\epsilon
_{21}$ and $\epsilon_{31}$, the condition
$\Delta=0$ represents, via Eq.\,(\ref{eq:del}),
a curve in the $({\sf O}\epsilon_{21}\epsilon_
{31})$ plane.   The
transition is forbidden for all values of
the axis ratios, which make a negative
discriminant, and then imaginary solutions.

The solutions of Eq.\,(\ref{eq:tre2a}) are:
\begin{lefteqnarray}
\label{eq:sol2a}
&& {\cal E}_{rot}^\prime=b\mp\left\{
b^2-\frac{\beta_M^5}{\beta_J^2\beta_E}\left(
\frac{{\cal S}}{{\cal S}^\prime}\right)^2\frac
{{\cal R}^\prime}{{\cal R}}{\cal E}_{rot}
\left[\frac{2\zeta-1}{2\zeta}-\frac{\zeta-1}
\zeta{\cal E}_{rot}\right]\right\}^{1/2}
~~;
\end{lefteqnarray}
and the combination of Eqs.\,(\ref{eq:teada})
and (\ref{eq:sol2a}) yields:
\begin{lefteqnarray}
\label{eq:rai2}
&& \frac{a_1}{a_1^\prime}=\frac{\beta_J^2}{\beta_M^3}
\frac{{\cal S}^\prime}{{\cal S}}\frac{{\cal R}}
{{\cal R}^\prime}\frac1{{\cal E}_{rot}}\left\{
b\mp\left[b^2-\frac{\beta_M^5}{\beta_J^2\beta_E}
\left(\frac{{\cal S}}{{\cal S}^\prime}\right)^2
\frac{{\cal R}^\prime}{{\cal R}}{\cal E}_{rot}
\left(\frac{2\zeta-1}{2\zeta}
-\frac{\zeta-1}\zeta{\cal E}_{rot}\right)\right]
^{1/2}\right\}~~; \nonumber \\
&&
\end{lefteqnarray}
where the upper and lower sign correspond
to the upper and lower sign in Eq.\,(\ref
{eq:sol2a}).   It is worth noticing that
the rotation parameter, ${\cal E}_{rot}^
\prime$, and the axis ratio, $a_1/a_1^
\prime$, are left unchanged for different
departures from mass, angular momentum,
and energy conservation, provided the
ratios:
\begin{equation}
\label{eq:betea}
\beta_{\cal{E}}=\frac{\beta_M^5}{\beta_J^2\beta_E}
~~;\qquad\beta_a= \frac{\beta_J^2}{\beta_M^3}~~;
\end{equation}
do not vary.

The explicit expression of Eq.\,(\ref
{eq:RSz}) defines a curve
in the $({\sf O}\epsilon_{21}\epsilon_{31})$
plane, where the axis ratios of 
the final configuration must necessarily lie.   
The intersection between the curves,
represented by Eqs.\,(\ref{eq:RSz}) and
(\ref{eq:del}), yields a point where
the axis ratios of the limiting, final 
configuration lies, related to a null
discriminant.   Final configurations with
axis ratios, related to a negative
discriminant, cannot occur.

\subsection{The limiting case $c\to0$}
\label{czero}

The limiting case $c\to0$, via Eqs.\,(\ref
{seq:tre2}), corresponds to ${\cal E}_{rot}^
\prime=0$ i.e. nonrotating initial
configurations, and to ${\cal E}_{rot}^\prime
=2b=1-{\cal E}_{osc}^\prime-{\cal E}_{pec}^
\prime$, i.e. unbound configurations.  
Further inspection to Eq.\,(\ref{eq:tre2c})
discloses that the right hand-side member
may tend to zero in a twofold manner, through
either the ratio ${\cal E}_{rot}/\beta_J^2$
or the difference within square brakets.

With regard to the former alternative, the
combination of Eqs.\,(\ref{eq:ErEb}), (\ref
{eq:teadb}),  and (\ref{eq:tra}) yields:
\begin{lefteqnarray}
\label{eq:sol0}
&& \frac{{\cal E}_{rot}}{\beta_J^2}=\frac{{\cal
R}}{{\cal S}}\frac{(J^\prime)^2}{GM^3a_1}~~; 
\end{lefteqnarray}
which tends to zero for nonrotating initial
configurations and/or infinitely extended
final configurations.

With regard to the latter alternative, the
difference within square brakets in 
Eq.\,(\ref{eq:tre2c}) tends to zero provided
the rotation parameter, ${\cal E}_{rot}$,
tends to the value:
\begin{equation}
\label{eq:erzi}
{\cal E}_{rot}=\frac12\frac{1-2\zeta}{1-\zeta}
~~;\qquad0\le\zeta\le\frac12~~;
\end{equation}
where the condition on the virial index, 
$\zeta$, is necessary to avoid negative values
of the rotation parameter in the case under
discussion.   The combination of Eqs.\,(\ref
{eq:virte}), (\ref{seq:zirf}),  and (\ref
{eq:erzi}) yields a second-degree equation
involving ${\cal E}_{rot}$ as unknown
and ${{\cal E}_{res}}$ as parameter, where
${\cal E}_{res}={\cal E}_{osc}+{\cal E}_{pec}$
in the case under discussion.   The solutions
of the above mentioned equation are:
\begin{equation}
\label{eq:qrot} 
{\cal E}_{rot}=\frac14~~;\qquad{\cal E}_{rot}=
\frac12-{\cal E}_{res}~~;
\end{equation}
and the substitution into Eq.\,(\ref{eq:erzi})
yields:
\begin{equation}
\label{eq:qres} 
\zeta=\frac13~~;\qquad\zeta=\frac{2{\cal E}_
{res}}{1+2{\cal E}_{res}}~~;
\end{equation}
where the former solution, via Eqs.\,(\ref
{eq:virte}) and (\ref{seq:zirf}), corresponds
to ${\cal E}_{res}=3/4$, and then to an
unbound i.e. infinitely extended final
configuration, as ${\cal E}_{rot}+{\cal E}_
{res}=1$.   On the other hand, the latter
solution clearly represents a relaxed
configuration, as ${\cal E}_{rot}+{\cal E}_
{res}=0.5$, which would necessarily imply
$\zeta=1$,
contrary to what has been found.   Then the
limiting case $c\to0$ is attained by making
the ratio ${\cal E}_{rot}/\beta_J^2$ tend
to zero.

\subsection{Virialized, final configurations}
\label{tranv}

In the special case of relaxed, final 
configurations, $\zeta=1$, virial equilibrium 
coincides with dynamical (or hydrostatic)
equilibrium.
Accordingly, Eq.\,(\ref{eq:tre2c}) 
translates into:
\begin{lefteqnarray}
\label{eq:bcc}
&& c=\frac12\frac{\beta_M^5}
{\beta_J^2\beta_E}\left(\frac{{\cal S}}
{{\cal S}^\prime}\right)^2\frac{{\cal R}^\prime}
{{\cal R}}{\cal E}_{rot}~~;
\end{lefteqnarray}
similarly, Eqs.\,(\ref{eq:sol2a}) and 
(\ref{eq:rai2}) reduce to:
\begin{lefteqnarray}
\label{eq:solv2}
&& {\cal E}_{rot}^\prime=
b\mp\left[b^2-\frac12\frac{\beta_M^5}{\beta_J^2
\beta_E}\left(\frac{{\cal S}}
{{\cal S}^\prime}\right)^2\frac{{\cal R}^\prime}
{{\cal R}}{\cal E}_{rot}\right]^{1/2}~~; \\
\label{eq:raiv2}
&& \frac{a_1}{a_1^\prime}=\frac{\beta_J^2}
{\beta_M^3}\frac{{\cal S}^\prime}{{\cal S}}\frac
{{\cal R}}{{\cal R}^\prime}\frac1{{\cal E}_{rot}}
\left\{b\mp\left[b^2-\frac12\frac{\beta_M^5}
{\beta_J^2\beta_E}\left(\frac{{\cal S}}
{{\cal S}^\prime}\right)^2\frac{{\cal R}^\prime}
{{\cal R}}{\cal E}_{rot}\right]^{1/2}\right\}~~;
\end{lefteqnarray}
where the upper and lower sign must correspond 
in both equations.


\subsection{Nonrotating, initial configurations}
\label{tranr}

In the special case of initial configurations
with vanishing angular momentum, $J^\prime\to0$,
the parameter, $\beta_J^{-1}$, has necessarily
to remain finite.   Accordingly, the combination
of Eqs.\,(\ref{eq:ErEb}), (\ref{eq:teadb}), and
(\ref{eq:tra}) yields:
\begin{lefteqnarray}
\label{eq:bJ1Er}
&& \beta_J^2{\cal E}_{rot}^\prime=\frac{{\cal R}^\prime}
{{\cal S}^\prime}\frac{J^2}{G(M^\prime)^3a_1^\prime}~~; \\
\label{eq:bJ2Er}
&& \beta_J^2({\cal E}_{rot}^\prime)^2=\left(\frac{{\cal R}^
\prime}{{\cal S}^\prime}\right)^2\frac{J^2(J^\prime)^2}
{G^2(M^\prime)^6(a_1^\prime)^2}~~; 
\end{lefteqnarray}
where the product on the left-hand side of
Eq.\,(\ref{eq:bJ2Er}) is infinitesimal of
higher order with respect to its counterpart
related to Eq.\,(\ref{eq:bJ1Er}).   Then 
Eq.\,(\ref{eq:tre2a}) reduces to a first-degree 
equation, and Eqs.\,(\ref{eq:sol2a}) and (\ref
{eq:rai2}) reduce to:
\begin{lefteqnarray}
\label{eq:rar2}
&& \beta_J^2{\cal E}_{rot}^\prime=\frac12\frac
{\beta_M^5}{\beta_E}\left(\frac{{\cal S}}{{\cal 
S}^\prime}\right)^2\frac{{\cal R}^\prime}{{\cal 
R}}{\cal E}_{rot}\left[\frac{2\zeta-1}{2\zeta}-
\frac{\zeta-1}\zeta{\cal E}_{rot}\right]\frac1b
~~; \\
\label{eq:rar0}
&& \frac{a_1}{a_1^\prime}=\frac12\frac{\beta_M^2}
{\beta_E}\frac{{\cal S}}{{\cal S}^\prime}
\left[\frac{2\zeta-1}{2\zeta}-\frac{\zeta-1}
\zeta{\cal E}_{rot}\right]\frac1b~~;
\end{lefteqnarray}
respectively.   It is worth noticing that
unbound configurations correspond to $b=0$,
via Eqs.\,(\ref{eq:teneb}) and (\ref
{eq:tre2b}).  

If, in addition, the rotation energy of the 
final configuration is negligible in respect 
of the self-potential energy, ${\cal E}_{rot}
\to0$, 
Eqs.\,(\ref{eq:rar2}) and (\ref{eq:rar0}) 
reduce to:
\begin{lefteqnarray}
\label{eq:rar3}
&& \beta_J^2{\cal E}_{rot}^\prime
=\frac12\frac{\beta_M^5}{\beta_E}\left(
\frac{{\cal S}}{{\cal S}^\prime}\right)^2\frac
{{\cal R}^\prime}{{\cal R}}{\cal E}_{rot}\frac
{2\zeta-1}{2\zeta}\frac1b~~; \\
\label{eq:rar1}
&& \frac{a_1}{a_1^\prime}=\frac12\frac{\beta_M^2}
{\beta_E}\frac{{\cal S}}{{\cal S}^\prime}\frac
{2\zeta-1}{2\zeta}\frac1b~~;
\end{lefteqnarray}
where both sides of Eq.\,(\ref{eq:rar3})
have necessarily to be infinitesimal of
the same order.

In the special case of a relaxed, final 
configuration, which necessarily implies
$\zeta=1$ according to
Eqs.\,(\ref{eq:virte}) and (\ref{seq:zirf}),
Eqs.\,(\ref{eq:solv2}) 
and (\ref{eq:raiv2}) reduce to Eqs.\,(\ref
{eq:rar2}) and (\ref{eq:rar0}), respectively, 
particularized
to the special case, $\zeta=1$.   The result is:
\begin{lefteqnarray}
\label{eq:rarv1}
&& \beta_J^2{\cal E}_{rot}^\prime=\frac14\frac
{\beta_M^5}{\beta_E}\left(\frac{{\cal S}}{{\cal 
S}^\prime}\right)^2\frac{{\cal R}^\prime}{{\cal 
R}}{\cal E}_{rot}\frac1b~~; \\
\label{eq:rarv0}
&& \frac{a_1}{a_1^\prime}=\frac14
\frac{\beta_M^2}{\beta_E}\frac{{\cal S}}
{{\cal S}^\prime}\frac1b~~;
\end{lefteqnarray}
where 
the trend is the same as shown above.   

\subsection{Application to dark matter haloes}
\label{trana}

Dark matter haloes make a natural
application of the procedure outlined above,
as they are the counterpart of
elliptical galaxies passing from
baryonic to dark matter universes.
Aiming to see nothing but the method
at work, our attention shall be 
restricted here to (i) CDM cosmological 
models and (ii) (non baryonic) dark 
matter, which is equivalent to assume 
a homeoidally striated density profile 
also for baryonic matter where, in 
addition, the isopycnic surfaces are
similar and similarly placed with 
respect to the boundary (e.g.,
CM03).

The current procedure has a main
advantage on numerical simulations,
that the initial configuration may
be taken as an overdensity region
at recombination epoch.  In this
case, the dominant energies are 
expected to be the potential
and the kinetic one, the latter 
related to systematic, expansion
motions, and the mass distribution
may safely be thought of as a 
homogeneous sphere.   Computations
usually start much later, owing to
their high cost and time machine.
The choice of recombination epoch
as starting time makes the initial
configuration virtually independent
of the cosmological model, at least
for sufficiently large, present-day 
density parameters $(\Omega_0\geq0.1)$, 
where the related universe may be
considered flat to a good extent
i.e. $\vert1-\Omega_{rec}^{-1}\vert
\ll1$ (e.g., Zeldovich \& Novikov
1982, Chap.\,III, \S 4; Caimmi \&
Marmo 2004).

Let us represent overdense regions
as made of two subsystems with
coinciding volume, namely (i) the
unperturbed, cosmological fluid,
with density equal to the mean
density of the universe, $\rho_h$,
and (ii) the density perturbation,
with mean density equal to the 
difference between the overdensity
and the mean density of the universe,
$\delta\bar{\rho}=\bar{\rho}-\rho_h$.
At recombination epoch, overdense
regions may safely be thought of as
homogeneous and spherical.   This is
why the overdensity index, $\bar
{\delta}_{rec}=\delta\bar{\rho}/\rho_h$,
cannot exceed (in absolute value)
a few percents at that time (e.g.,
Gunn 1987).   In addition, tidal
distorsions from sphericity are
owing to neighbouring density
perturbations, while tidal
distorsions towards sphericity 
are owing to the unperturbed,
cosmological fluid.

In general, the potential-energy
tensor of an overdensity region
is the sum of three contributions,
namely: (i) the self-energy tensor
of the overdense region, (ii) the
tidal-energy tensor induced by the
unperturbed, cosmological fluid
outside the overdense region, and
(iii) the tidal-energy tensor 
induced by neighbouring density
perturbations.   For reasons
mentioned above, the last two
contributions may safely be 
neglected in the case under
discussion.   Accordingly, the
combination of Eqs.\,(\ref{eq:Espq}),
(\ref{eq:Bspq}), and (\ref{eq:Spq}); 
(\ref{seq:ErJ}) and (\ref{eq:Rpq}); 
yields:
\begin{lefteqnarray}
\label{eq:Srec}
&& {\cal S}_{pq}^\prime=\frac15\delta_
{pq}~~;\quad{\cal S}^\prime=\frac
35~~; \\
\label{eq:Rrec}
&& {\cal R}_{pq}^\prime=\frac58
\delta_{pq}(1-\delta_{p3})~~;
\quad{\cal R}^\prime=\frac54~~;
\end{lefteqnarray}
respectively, as $\nu_{sel}=3/10$
and $A_1=A_2=A_3=2/3$ in the case 
under discussion of a homogeneous 
sphere; and $\eta_{anm}=2\eta_{rot}
=1$, $\nu_{anm}=\nu_{rot}=\nu_{inr}
=1/5$ for a homogeneous, rigidly
rotating sphere.

The kinetic-energy tensor of systematic,
expansion motions, is:
\begin{equation}
\label{eq:Eros}
(E_{osc}^\prime)_{pq}=\frac1{10}
\delta_{pq}M^\prime\dot{a}^{\prime2}~~;
\quad E_{osc}^\prime=\frac3{10}M^
\prime\dot{a}^{\prime2}~~;
\end{equation}
in the case under discussion of a
homogeneous sphere.   The further
assumption that, at recombination
epoch, the overdense region expands
at the same rate as the unperturbed,
cosmological fluid, implies the
validity of the relation:
\begin{equation}
\label{eq:apre}
\dot{a}^\prime=(\dot{a}_h)_{rec}
=H_{rec}(a_h)_{rec}=H_{rec}a^\prime~~;
\end{equation}
where $a_h$ is the scale factor of the
universe, $H$ is the Hubble parameter, 
and the index $rec$ denotes the 
recombination epoch.

At sufficiently early times i.e. high 
redshifts, in particular at the
recombination epoch ($z_{rec}\sim
1400$), the following
relations hold (e.g., Zeldovich \& 
Novikov 1982, Chap.\,III, \S 4):
\begin{lefteqnarray}
\label{eq:omz}
&& \Omega^{-1}(z)=1+\frac1{1+z}\frac
{1-\Omega_0}{\Omega_0}~~;\quad\Omega
_0\geq0.1~~; \\
\label{eq:rhoh}
&& \rho_h(z)=\frac3{8\pi}\frac{H_0^2}
G(1+z)^3~~;\quad\vert1-\Omega^{-1}
\vert\ll1~~; \\
\label{eq:Hz}
&& H(z)=\frac23\frac1t=H_0(1+z)^{3/2}
~~;\quad\vert1-\Omega^{-1}\vert\ll1~~;
\end{lefteqnarray}
where $\Omega$ is the density parameter
(to be not confused with the angular
velocity), $z$ the redshift, $t$ the
time, and the index 0 denotes the
present-day epoch.   The definition of
the density parameter reads:
\begin{equation}
\label{eq:pdc}
\Omega=\frac{\rho_h}{(\rho_h)_{crit}}~~;
\qquad(\rho_h)_{crit}=\frac{3H^2}{8\pi G}~~;
\end{equation}
where $(\rho_h)_{crit}$ is the critical
density i.e. the density of the Hubble
flow in an Einstein-de Sitter universe.

By use of the definition of mean 
density inside an overdense region,
$\bar{\rho}=\rho_h+\delta\bar{\rho}
=\rho_h(1+\bar{\delta}
)=3M_h(1+\bar{\delta}
)/(4\pi
a_h^3)$, and Eqs.\,(\ref{eq:apre}), 
(\ref{eq:rhoh}), and (\ref{eq:Hz}), 
the following relations are derived 
after some algebra:
\begin{lefteqnarray}
\label{eq:ari}
&& a^\prime=\left[\frac{2GM^\prime}
{H_0^2(1+\bar{\delta}_{rec})}\right]
^{1/3}\frac1{1+z_{rec}}~~; \\
\label{eq:vri}
&& \dot{a}^\prime=H_0(1+z_{rec})^{3/2}
a^\prime=\left(\frac{2GM^\prime H_0}
{1+\bar{\delta}_{rec}}\right)^{1/3}
(1+z_{rec})^{1/2}~~;
\end{lefteqnarray}
and the combination of Eqs.\,(\ref
{eq:Eros}) and (\ref{eq:apre}) reads:
\begin{lefteqnarray}
\label{eq:Eo}
&& (E_{osc}^\prime)_{pq}=\frac1{10}\delta_
{pq}H_{rec}^2M^\prime a^{\prime2}~~;\quad 
E_{osc}^\prime=\frac3{10}H_{rec}^2M^\prime 
a^{\prime2}~~;
\end{lefteqnarray}
in the case under consideration of a
homogeneous sphere expanding at the 
same rate as the universe.   The
combination of Eqs.\,(\ref{eq:Spq}),
(\ref{eq:tEse}), (\ref{eq:teneb}), 
(\ref{eq:pdc}) and (\ref{eq:Eo}) yields:
\begin{equation}
\label{eq:Erore}
{\cal E}_{osc}^\prime=\left[\Omega_{rec}(1+\bar
{\delta}_{rec})\right]^{-1}~~;
\end{equation}
which has necessarily to be less than
unity for bound density perturbations.

Accordingly, the sum of initial systematic
rotational and peculiar energy, $E_{rot}^
\prime+E_{pec}^\prime $, cannot exceed 
$-E_{pot}^\prime-E_{osc}^\prime$ for bound 
configurations, i.e. ${\cal E}_{rot}^\prime+
{\cal E}_{pec}^\prime<1-{\cal E}_{osc}^\prime$.   
Keeping
in mind that $\Omega_{rec}$ may safely be
put equal to unity and the overdensity
index, $\bar{\delta}_{rec}$, amounts to a few
percent, it can be seen from Eq.\,(\ref
{eq:Erore}) that the sum of parameters, 
${\cal E}_{rot}^\prime+{\cal E}_{pec}^
\prime$, cannot 
exceed a few percent, or in other words 
the sum of initial rotational and peculiar
energy is two orders of 
magnitude lower than the initial expansion
energy.   Then the knowledge of the initial 
rotational and peculiar velocity field is 
not essential, and for this reason an 
initial rigid rotation is assumed, together
with a null, initial, peculiar velocity 
field.

With regard to the final configuration,
the simplest case of mass conservation,
$M=M^\prime$ or $\beta_M=1$, and energy
conservation, $E=E^\prime$ or $\beta_E=1$,
makes a valid approximation for isolated,
overdense regions.   On the other hand,
merger accretion necessarily implies 
$\beta_M>1$ and $\beta_E>1$.   In addition, 
(complete) virialization, $E_{sel}+2(E_{rot}+E_
{res})=0$, may safely be assumed in dealing
with density perturbations of galactic size,
while it must be released for larger dimensions 
i.e. clusters of galaxies, where the inner
region is relaxed but the outer shells are
still infalling.

For representing density profiles of virialized,
dark matter haloes, expressed by Eqs.\,(\ref
{seq:profg}), two alternatives will be exploited.
One, first proposed by Navarro et al. (1995; 
1996; 1997), reads:
\begin{equation}
\label{eq:NFW}
f(\xi)=\frac4{\xi(1+\xi)^2}~~;
\end{equation}
and will be quoted in the following as the NFW
density profile.    One other, first proposed
by Moore et al. (1998; 1999), reads:
\begin{equation}
\label{eq:MOA}
f(\xi)=\frac2{\xi^{3/2}(1+\xi^{3/2})}~~;
\end{equation}
and will be quoted in the following as the MOA
density profile.   In both cases, the truncated
scaled radius is determined by fitting the results
of high-resolution simulations in a CDM cosmological
model (Fukushige \& Makino 2001) to the selected
density profile.    For more details, see CM03; 
Caimmi \& Marmo (2004).   In addition, both rigid
rotation (triaxial configurations) and constant
rotational velocity on the equatorial plane 
(axisymmetric configurations) are assumed.    In
the latter alternative, both rigid ($\eta_{anm}=2
\eta_{rot}=1$) and differential (to ensure constant
rotational velocity everywhere: $\eta_{anm}=3\pi/
8$, $\eta_{rot}=3/4$) rotation of the generic,
isopycnic surface, are considered.   The values of 
the truncated, scaled radius, and some profile 
parameters, expressed in Sect.\,\ref{gente} and 
particularized to both NFW and MOA density profiles, 
are listed in Tab.\,\ref{t:profa}. 
\begin{table}
\begin{tabular}{lll}
\hline
\hline
\multicolumn{1}{c}{parameter} &
\multicolumn{1}{c}{NFW} &
\multicolumn{1}{c}{MOA} \\
\hline
$\Xi$ & \phantom{1}9.20678 & \phantom{1}3.80693 \\
$\nu_{mas}$ & 17.05231 & \phantom{1}8.52616 \\
$\nu_{inr}$ & \phantom{1}0.0554130 & \phantom{1}0.0892287 \\
$\nu_{sel}$ & \phantom{1}0.610045 & \phantom{1}0.561202 \\
$\nu_{anm}$ & \phantom{1}0.139180 & \phantom{1}0.146741 \\
$\nu_{rot}$ & \phantom{1}0.333333 & \phantom{1}0.333333 \\
$\nu_{ram}$ & 17.20783 & 15.48009 \\
\hline\hline
\end{tabular}
\caption{Values of the truncated, scaling radius,
$\Xi$, and some profile parameters, expressed in
Sect.\,\ref{gente} and particularized to both 
NFW and MOA density profiles, after comparison
with the results of high-resolution simulations
(Fukushige \& Makino 2001).
See text for more details.   The last three 
parameters, related to the rotational velocity
field, have been calculated in the special case
of constant rotational velocity on the equatorial
plane (axisymmetric configurations) related to 
either rigid ($\eta_{anm}=2
\eta_{rot}=1$) or differential (to ensure constant
rotational velocity everywhere: $\eta_{anm}=3\pi/
8$, $\eta_{rot}=3/4$) rotation of the generic,
isopycnic surface.   In the special case of rigidly 
rotating, triaxial configurations, $\nu_{anm}=
\nu_{rot}=\nu_{inr}$, $\nu_{ram}=1/\nu_{inr}$.}
\label{t:profa}
\end{table}

At this stage, the calculations are allowed to start,
following the procedure outlined below.
\begin{description}
\item[\rm{(i)}\hspace{0.2mm}] Define the
values of the input parameters: the axis
ratios, $\epsilon_{21}^\prime$ and $\epsilon
_{31}^\prime$, the profile parameters, $\nu_
{sel}^\prime$ and $\nu_{ram}^\prime$, the 
overdensity index, $\bar{\delta}_{rec}$, the 
degree of departure from mass conservation, 
$\beta_M$, energy conservation, $\beta_E$, 
angular momentum conservation, $\beta_J$, 
and the virial index, $\zeta$, the last in 
connection with the final configuration.
\item[\rm{(ii)}~~] Select a density profile
for the final configuration, and calculate
the profile parameters, $\nu_{mas}$, $\nu_
{inr}$, $\nu_{sel}$.
\item[\rm{(iii)}~] Select a velocity profile  
(of the kind considered in Sect.\,\ref{gente})
for the final configuration, and calculate the
profile parameters, $\nu_{anm}$, $\nu_{rot}$,
$\nu_{ram}$, and the shape parameters, $\eta_
{anm}$, $\eta_{rot}$. 
\item[\rm{(iv)}\hspace{1.7mm}] Select a 
sequence of homeoidally striated Jacobi
ellipsoids (which represent final 
configurations), where the generalized and
effective anisotropy parameters, $\zeta_
p$ and $\tilde{\zeta}_{pp}$, are kept constant.
\item[\rm{(v)}\hspace{2.5mm}]  Select a
value of the axis ratio, $\epsilon_{21}$,
of a configuration on the sequence under
consideration.
\item[\rm{(vi)}\hspace{1.7mm}] Calculate
the value of the remaining axis ratio,
$\epsilon_{31}$. 
\item[\rm{(vii)}\hspace{1mm}] Calculate the value 
of the rotation parameters, $h$, ${\cal E}_{rot}$,
$\upsilon$, $\lambda$, and $\chi_v$,
defined 
by Eqs.\,(\ref{eq:ch}), (\ref{eq:crErs}), 
(\ref{eq:cupsh}), (\ref{eq:clam3}), and 
(\ref{eq:cvrope}), respectively.
\item[\rm{(viii)}] Calculate the value of the
ratios, ${\cal E}_{rot}^\prime/(1-{\cal E}_
{osc}^\prime)$, and  $a_1/a_1^\prime$, defined 
by Eqs.\,(\ref{seq:tre2}), (\ref{eq:sol2a}),
(\ref{eq:rai2}), and (\ref{eq:Erore}).
\item[\rm{(ix)}\hspace{1.7mm}] Select another
configuration on the sequence under 
consideration, and return to (v), unless the 
whole sequence is covered.
\item[\rm{(x)}\hspace{2.5mm}] Select another 
sequence of homeoidally striated Jacobi
ellipsoids, where the generalized and
effective anisotropy parameters, $\zeta_
{pp}$ and $\tilde{\zeta}_{pp}$, are kept constant,
and return to (v).
\item[\rm{(xi)}\hspace{1.9mm}] Select another
velocity profile for the final configuration,
and return to (iv), unless the range of interest
is covered.
\item[\rm{(xii)}\hspace{0.9mm}] Select another 
density profile
for the final configuration, and return to (iii),
unless the range of interest is covered.
\item[\rm{(xiv)}\hspace{.5mm}] Select another 
value of the virial index, $\zeta$
(in connection with the final 
configuration),  and return to (ii), 
unless the range of interest is covered.
\end{description}

\subsection{Results and discussion}
\label{disc}

Our attention shall be limited to dark matter
haloes hosting giant galaxies, with masses of
about $10^{12}{\rm m}_\odot$.   Accordingly, we
assume $\bar{\delta}_{rec}=0.015$ as a typical
overdensity index of the initial configuration,
and a virialized final configuration i.e. 
$\zeta=1$.   In addition, the following range
of parameters, $\beta_{{\cal E}}$ and $\beta_a$,
is considered:
\begin{equation}
\label{eq:ibea}
\frac1{100}\ge\beta_{{\cal E}}\ge\frac1{14400}~~;
\qquad100\le\beta_a\le14400~~;
\end{equation}
which includes the case $\beta_M=\beta_E=1$,
allowing $\beta_J$ lie within the range:
\begin{equation}
\label{eq:ibj}
10\le\beta_J\le120~~;
\end{equation}
owing to Eq.\,(\ref{eq:betea}).   For sake of
simplicity, in what follows the last choice
of $(\beta_M, \beta_E, \beta_J)$ will be
adopted, but keeping in mind that it is
equivalent to any other choice satisfying
Eqs.\,(\ref{eq:ibea}), for what will
be analysed and discussed.

In particular, three sequences of density
profiles with constant anisotropy parameters,
$\tilde{\zeta}_{pp}$, are considered, starting
from a nonrotating, axisymmetric configuration
equal to an oblate $(\epsilon_{31}=0.5;~\tilde
{\zeta}_{11}=0.3910;~\tilde{\zeta}_{33}=0.2180)$,
round $(\epsilon_{31}=1;~\tilde{\zeta}_{11}=\tilde
{\zeta}_{33}=1/3)$, and prolate $(\epsilon_{31}=2;~
\tilde{\zeta}_{11}=0.2717;~\tilde{\zeta}_{33}=0.4566)$
spheroid, respectively.

\begin{figure}
   \centering
\resizebox{\hsize}{!}{\includegraphics{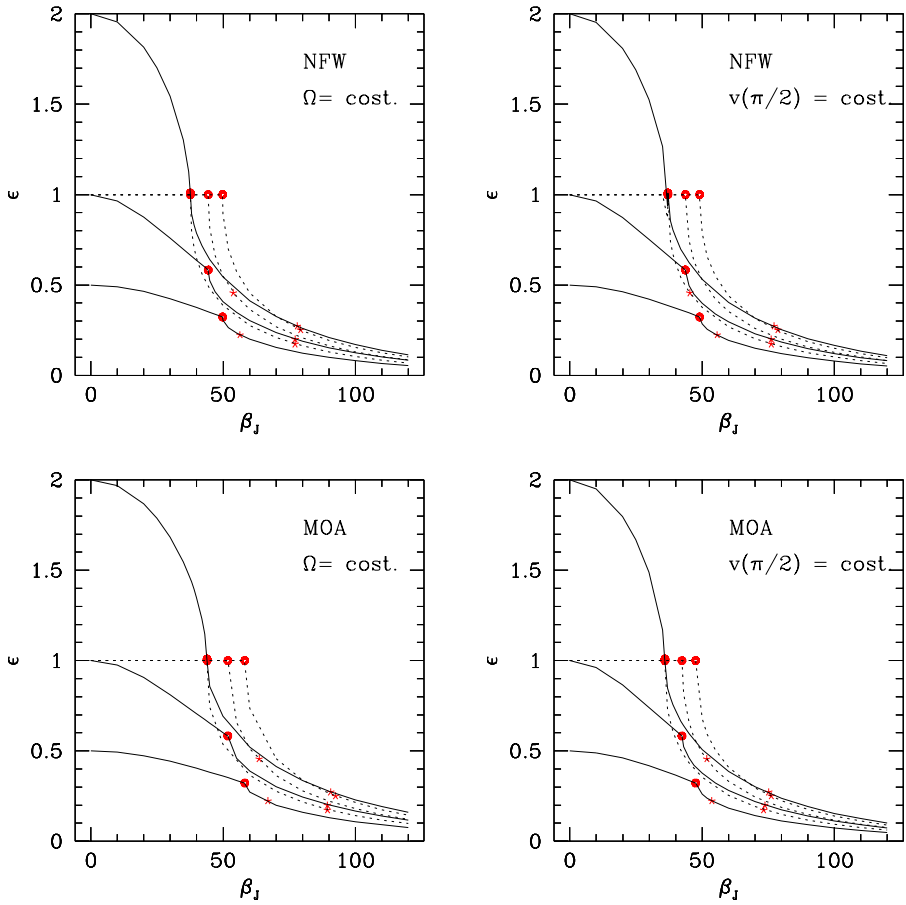}} 
\caption{The axis ratios of the limiting configuration
as a function of the parameter, $\beta_J$, for sequences
of virialized density profiles with constant anisotropy 
parameters, starting from axisymmetric, nonrotating  
configurations, defined as: O $(\epsilon_{31}=0.5;~\tilde
{\zeta}_{11}=0.3910;~\tilde{\zeta}_{33}=0.2180)$; S
$(\epsilon_{31}=1;~\tilde{\zeta}_{11}=\tilde{\zeta}_{33}=1/3)$; 
P $(\epsilon_{31}=2;~\tilde{\zeta}_{11}=0.2717;~\tilde{\zeta}
_{33}=0.4566)$.   The polar and the equatorial axis ratio,
$\epsilon_{31}$ and $\epsilon_{21}$, are represented by
full and dashed curves, respectively.   Configurations
for which bifurcation from axisymmetric to triaxial
density profiles occurs, are marked by filled circles.
Pentagon skeletons mark configurations for which the
validity of a necessary condition for the occurrence of
centrifugal support at the ends of the major axis, is
first satisfied.   The density profile and the kind
of rotation are indicated in each panel.   Configurations
related to constant rotational velocity on the equatorial
plane, $v(\pi/2)=const$, and rigid rotation of isopycnic
surfaces, have no physical counterpart for triaxial
boundaries.}
\label{f:reli}    
\end{figure}

As in Thuan \& Gott (1975) original approach,
a limiting configuration exists, which makes
the discriminant, expressed by Eq.\,(\ref
{eq:del}), equal to zero.   More flattened
configurations are not allowed
for the related sequence of density profiles
and choice of $(\beta_M, \beta_E, \beta_J)$.
The dependence of the axis ratios of the
limiting configuration on the parameter,
$\beta_J$, related to the three sequences
mentioned above, is represented in Fig.\,\ref
{f:reli} for NFW and MOA virialized density 
profiles, characterized by either rigid rotation 
or constant rotational velocity on the equatorial
plane and rigid rotation of isopycnic surfaces.
Configurations
for which bifurcation from axisymmetric to triaxial
density profiles occurs, are marked by filled circles.
Pentagon skeletons mark configurations for which the
validity of a necessary condition for the occurrence of
centrifugal support at the ends of the major axis, 
according to Eq.\,(\ref{eq:inev}), is first satisfied.

As low values of $\beta_J$ correspond to bounder final
configurations and vice versa, the axis ratio
of the limiting configurations are strongly dependent
on the sequence under consideration for low values of
$\beta_J$ and vice versa.   In other words, a large
amount of rotational energy erases the memory of the
initial shape, different for different sequences.
A complete dissipation of
angular momentum, $\beta_J=0$, yields nonrotating
final configurations.   In addition, it can be seen
from Fig.\,\ref{f:reli} that a conservation of angular
momentum, $\beta_J=1$, implies final configurations
lying near the nonrotating limit.

The density profile and the kind of rotation are
indicated in each panel of Fig.\,\ref{f:reli}.
Configurations
related to constant rotational velocity on the equatorial
plane, $v(\pi/2)=const$, and rigid rotation of isopycnic
surfaces, have no physical counterpart for triaxial
boundaries, but they may be of mathematical
interest for comparison with solid-body rotating mass
distributions with equal density profile.   On the
other hand, a change in density profile affects the
curves to a smaller extent for configurations with
constant rotational velocity on the equatorial plane,
with respect to rigidly rotating configurations, as
expected.

From this point on, our attention shall be focused
on two special values of $\beta_J$, namely:
\begin{eqnarray}
\label{eq:bJ60}
&& \beta_J=60~~;\qquad0.20\appleq(\epsilon_{31})_{lim}
\appleq0.40~~;\qquad0.30\appleq(\epsilon_{21})_{lim}
\appleq0.50~~; \\
\label{eq:bJ120}
&& \beta_J=120~~;\qquad0.05\appleq(\epsilon_{31})_{lim}
\appleq0.16~~;\qquad0.06\appleq(\epsilon_{21})_{lim}
\appleq0.15~~;
\end{eqnarray} 
in connection with NFW density profiles in rigid 
rotation.   An inspection to Fig.\,\ref{f:reli}
shows that the validity of a necessary condition 
for the occurrence of centrifugal support at the 
ends of the major axis, according to Eq.\,(\ref
{eq:inev}), may be violated.

\begin{figure}
   \centering
\resizebox{\hsize}{!}{\includegraphics{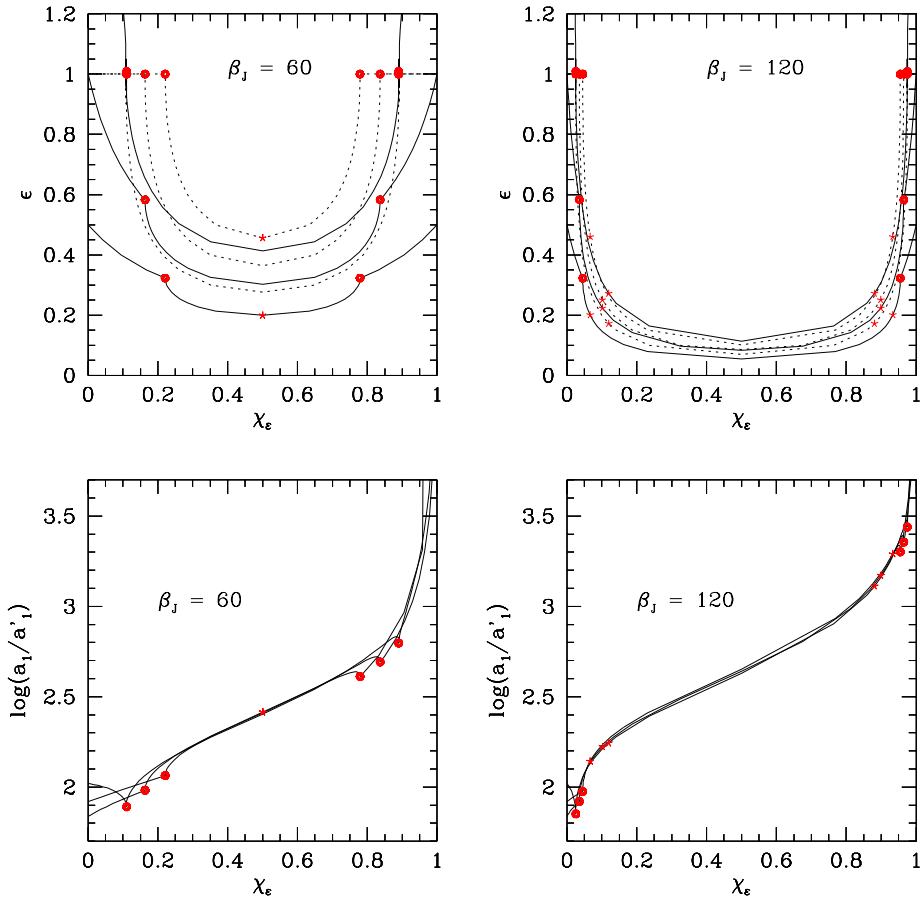}} 
\caption{The axis ratios of the final configuration,
$\epsilon_{31}$ and $\epsilon_{21}$ (upper panels)
and the decimal logarithm of the axis ratio, $a_1/
a_1^\prime$, of final to 
initial configuration (lower panels), as a function 
of the parameter, $\chi_{{\cal E}}={\cal E}_{rot}^ 
\prime/(1-{\cal E}_{osc}^\prime)$, for sequences
of virialized density profiles, O, S, and P, 
considered in Fig.\,\ref{f:reli}, top left panel,
related to $\beta_J=60$ (left panels) and $\beta_J= 
120$ (right panels).   The curves on the upper panels
are symmetric with respect to a vertical axis,
$\chi_{{\cal E}}=0.5$.   The point of minimum of
each curve of the upper panels corresponds to the
related, limiting value of the axis ratio, i.e. the
lower value allowed along the sequence under consideration.
Other captions as in Fig.\,\ref{f:reli}.   The limit,
$\chi_{{\cal E}}\to1$, is related to unbound configurations
$(E=0)$ which extend to infinite.}
\label{f:ESpe}    
\end{figure}

The axis ratios of the final configuration, 
$\epsilon_{31}$ and $\epsilon_{21}$ (upper
panels) and the axis ratio of final to initial
configuration, $a_1/a_1^\prime$ (lower panels),
as a function of the parameter, $\chi_{{\cal 
E}}={\cal E}_{rot}^\prime/(1-{\cal E}_{osc}^
\prime)$, are plotted in Fig.\,\ref{f:ESpe}
for three sequences of density profiles, O,
S, and P, considered earlier, and $\beta_J=
60$ (left panels) and $\beta_J=120$ (right
panels). 
Each curve on the upper panels
is symmetric with respect to a vertical
axis, $\chi_{{\cal E}}=0.5$, and is 
characterized by three extremal points, 
two maxima and one minimum.   The maxima 
are related to the starting configuration 
of the sequence, with finite
and infinite axes, respectively.   On the
other hand, the minimum corresponds to the
limiting configuration, which exhibits the
lowest values of the axis ratios, allowed
for the sequence under consideration.
Flat configurations occur only in the limit
$\beta_J\to+\infty$ unless centrifugal
support takes place at the ends of the 
major axis, from which two flows of
dark matter freely stream (see e.g.,
Jeans 1929, Chap.\,XIII, \S\,301, with
regard to gaseous nebulae).   In addition,
larger values of $\beta_J$ make different
curves closer each to the other, and vice
versa.   The limit, $\chi_{{\cal E}}\to1$,
corresponds to unbound configurations $(E
=0)$ which extend to infinite.

\begin{figure}
   \centering
\resizebox{\hsize}{!}{\includegraphics{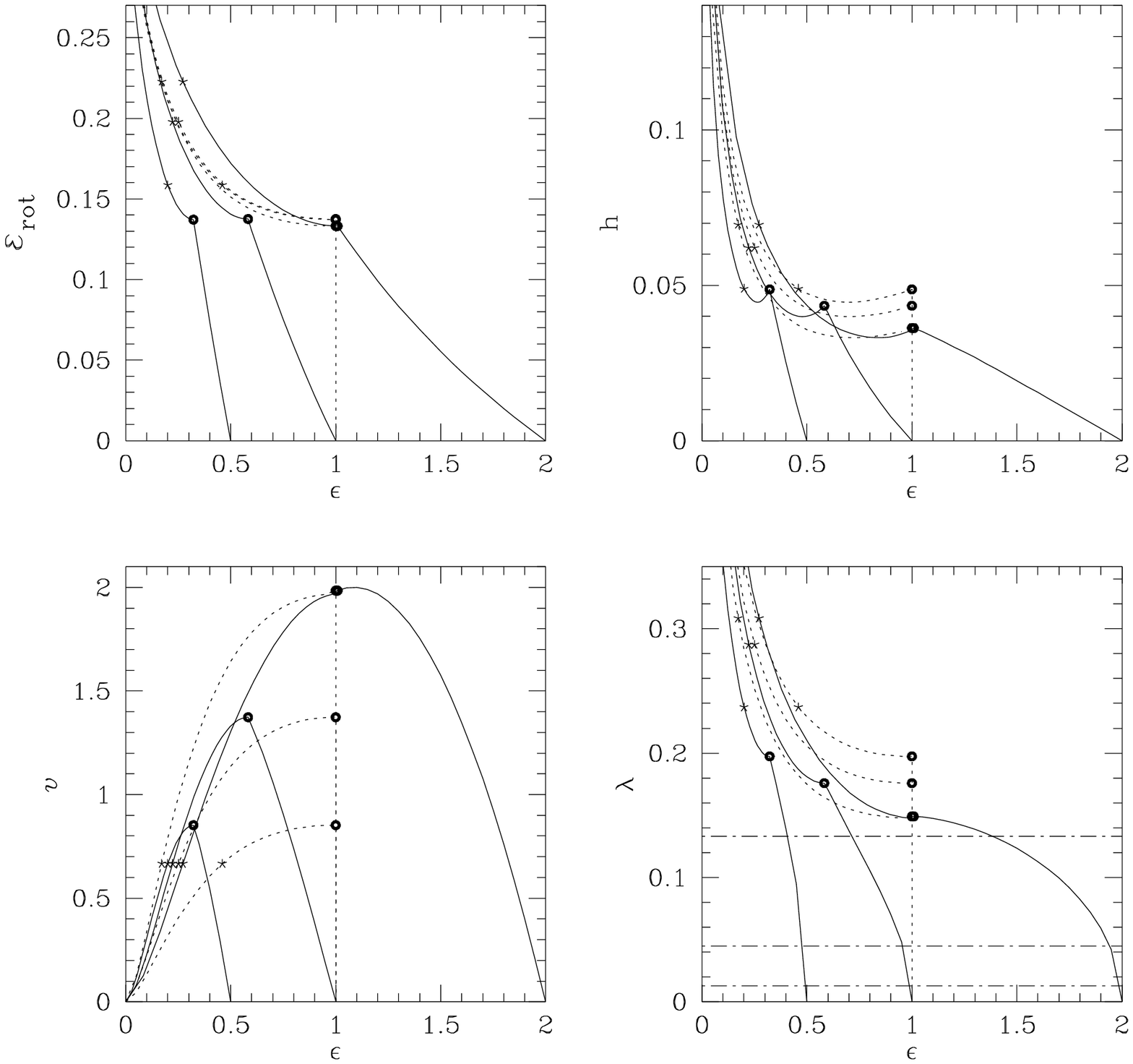}} 
\caption{The rotation parameters, ${\cal E}_{rot}$,
$h$, $\lambda$, and $\upsilon$, (from top 
left in clockwise sense), related to the
final configuration, as a function of the
axis ratios, $\epsilon_{31}$ (full curves)
and $\epsilon_{21}$ (dashed curves), for
sequences of density profiles, O, S, and
P, considered in Fig.\,\ref{f:reli}, top   
left panel.   The curves are independent
of the parameter, $\beta_J$, but the limiting
values of the axis ratios, related to $\beta_J$,
cannot be exceeded moving from the right
to the left along each curve.   The horizontal
lines correspond to a mean value, $\lambda=
0.0421$, and related variations within one
rms error, $\lambda^+=0.133$ and $\lambda^-
=0.0133$, consistent with results from
high-resolution numerical simulations.
Other captions as in Fig.\,\ref{f:reli}.}  
\label{f:prot}    
\end{figure}

\begin{figure}
   \centering
\resizebox{\hsize}{!}{\includegraphics{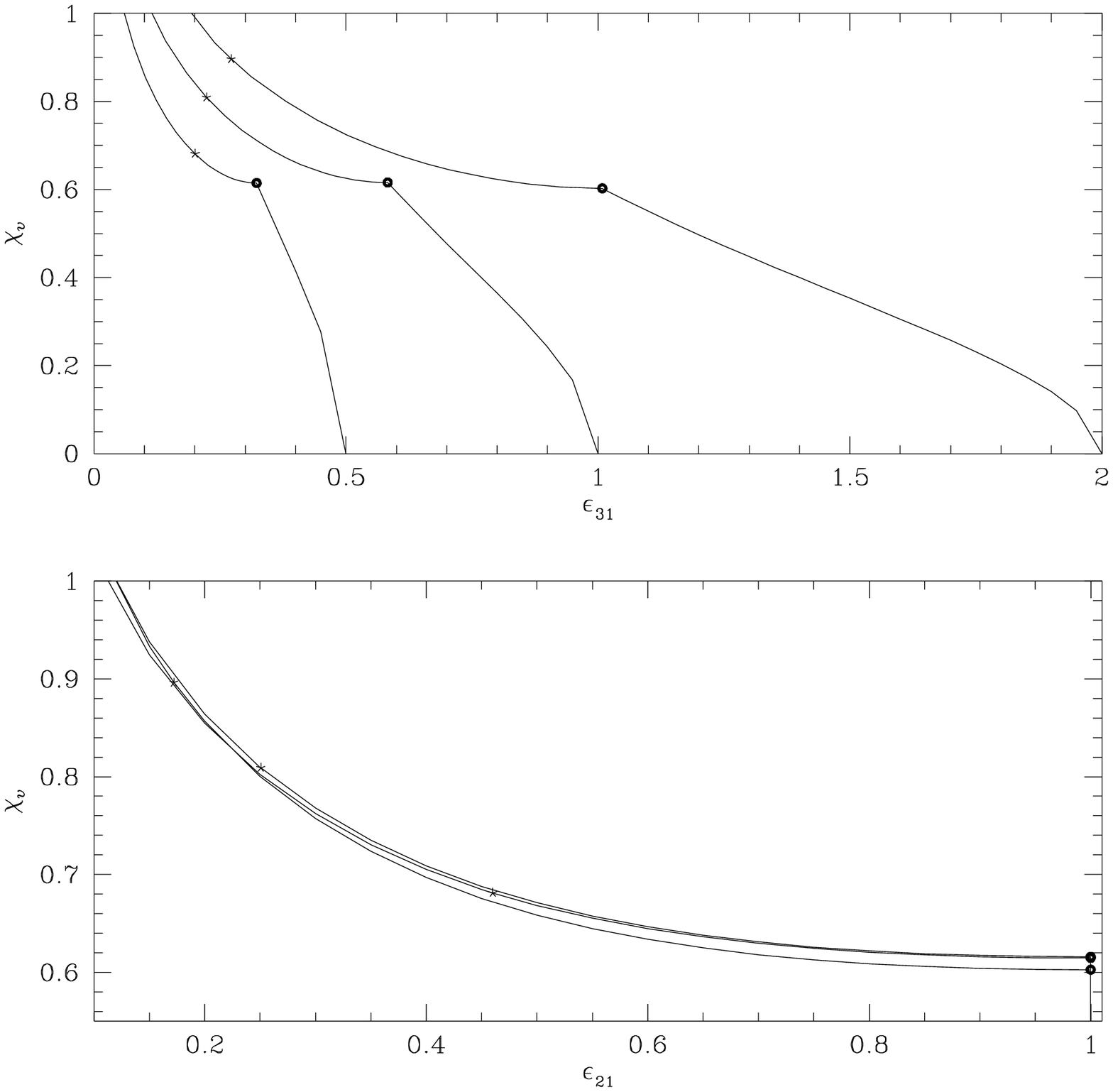}} 
\caption{The rotation parameter, $\chi_v$,
related to the
final configuration, as a function of the
axis ratios, $\epsilon_{31}$ (upper panel) and 
$\epsilon_{21}$ (lower panel), for
sequences of density profiles, O, S, and
P, considered in Fig.\,\ref{f:reli}, top   
left panel.   The curves are independent
of the parameter, $\beta_J$, but the limiting
values of the axis ratios, related to $\beta_J$,
cannot be exceeded moving from the right
to the left along each curve.   
Other captions as in Fig.\,\ref{f:reli}.}  
\label{f:rvrp}    
\end{figure}

The rotation parameters, ${\cal E}_{rot}$,
$h$, $\lambda$, and $\upsilon$,  (from top 
left in clockwise sense), and 
$\chi_v$,
related to the
final configuration, as a function of the
axis ratios, $\epsilon_{31}$ and $\epsilon
_{21}$, are plotted in Figs.\,\ref{f:prot}
and \ref{f:rvrp}, respectively, for three
sequences of density profiles, O, S, and
P, considered earlier.   
The curves are independent
of the parameter, $\beta_J$, but the limiting
values of the axis ratios, related to the
selected value of $\beta_J$,
cannot be exceeded moving from the right
to the left along each curve.   The horizontal
lines correspond to a mean value, $\lambda=
0.0421$, and related variations within one
rms error, $\lambda^+=0.133$ and $\lambda^-
=0.0133$, consistent with the results of
high-resolution numerical simulations
(e.g., Bullock et al. 2001; Vitvitska et al.
2002).

Leaving aside peculiar situations such as
the occurrence of a major merger near the
end of evolution, the main results of 
high-resolution simulations (e.g., Fukushige
\& Makino 2001; Klypin et al. 2001; Bullock
et al. 2001; Vitvitska et al. 2002) may be 
summarized as follows, to a good extent:
(i) a substantial fraction of the initial
density perturbation gets virialized; 
(ii) the density profile may
be considered as self-similar and universal; 
and (iii) the distribution
of the rotation parameter, $\lambda$, is
lognormal with mean value, $0.03\appleq
\lambda\appleq0.05$, and rms error, $0.5
\appleq\sigma_{\log\lambda}\appleq0.7$, in
particular $\lambda=0.0421$, $\sigma_{\log
\lambda}=0.500$.

Accordingly, simulated density profiles
with assigned mass, boundary, and rotation
parameter, may be compared with their
counterparts deduced from the current
model.   In particular, an inspection
of Figs.\,\ref{f:ESpe} and \ref{f:prot}
shows that, in the range $0.013\le
\lambda\le0.133$, dark matter haloes
are close to the starting point of
the selected sequence i.e. the
nonrotating configuration.   Then
the shape of dark matter haloes is
decided mainly by the amount of
anisotropy in residual velocity
distribution, while the contribution
of rotation has only a minor effect on
the meridional plane, and no effect on
the equatorial plane.   This is why
bifurcation points occur for larger
values of $\lambda$.
In other words, dark matter haloes
may exhibit any shape, provided it
is owing mainly to anisotropy in
residual velocity distribution
and only weakly to systematic
rotation.

The analysis of elongated configurations
which rotate around the major axis goes
beyond a purely academic interest.
Empirical evidence in favour of the
existence of galaxies with prolate
stellar structure, cut equatorially
by a gaseous lane, was found long
time ago (Bertola \& Galletta 1978).
The hosting dark haloes could also 
be prolate and spinning around the
symmetry axis.   A recent investigation
on polar ring galaxies has indicated
that the hosting dark haloes are most
likely elongated towards the polar
ring plane (Iodice et al. 2003), where
the rotation axis of the inner body lies.

\section{Conclusion}\label{conc}

The formulation of the tensor virial
equations has been generalized to unrelaxed
configurations, where virial equilibrium
does not coincide with dynamical (or
hydrostatic) equilibrium.   Special
classes of homeoidally striated ellipsoids
have been shown to exhibit similar
properties as Jacobi ellipsoids, and
to reduce to Jacobi ellipsoids in the
limiting situation of homogeneous
matter distribution, rigid rotation,
and isotropic residual velocity
distribution.   Accordingly, the
above mentioned density profiles
have been defined as homeoidally
striated, Jacobi ellipsoids.
Further investigation has been
devoted to the generation of sequences
of virial equilibrium configurations,
and sequences with constant rotation
and anisotropy
parameters have been studied with more
detail, including both flattened and
elongated triaxial (in particular, 
both oblate and prolate axisymmetric)
configurations, and the determination
of the related bifurcation points.
The explicit expression of a number 
of rotation parameters, used in
literature, and the ratio of rms
rotation to rms residual velocity,
have also been calculated.

Thuan \& Gott (1975) procedure has
been generalized to homeoidally
striated, Jacobi ellipsoids, and a
few limiting situations have also
been analysed.   An application
has been made to dark matter haloes,
and calculations have been performed
for masses $M\approx10^{12}{\rm m}_\odot$
i.e. haloes hosting giant galaxies.
The results have been shown to be
qualitatively similar to what was found
by Thuan \& Gott (1975), in particular
concerning the occurrence of limiting
axis ratios, below which no configuration
is allowed for the sequence under
consideration.   The dependence of the
limiting axis ratios on two parameters,
related to the degree of change in mass,
total energy, and angular momentum, has
been illustrated in connection with
NFW and MOA density profiles, rotating
with constant either angular or linear
(on the equatorial plane) velocity,
for three special sequences starting
from an oblate $(\epsilon_{31}=0.5)$,
round $(\epsilon_{31}=1)$, and
prolate $(\epsilon_{31}=2)$, nonrotating
configuration, respectively.

The dependence of the axis ratios, 
$\epsilon_{31}$ and $\epsilon_{21}$, on
the parameter, $\chi_{{\cal E}}={\cal 
E}_{rot}^ \prime/(1-{\cal E}_{osc}^
\prime)$, has been analysed in connection
with rigidly rotating, NFW density
profiles, for the three special sequences
mentioned above.   The same has been done 
for the rotation parameters. 
Within the range of the rotation parameter,
$\exp_{10}(\log\lambda\mp\sigma_{\log\lambda})=
\exp_{10}(-1.3761\mp0.500)$, consistent with
results from high-resolution numerical simulations,
it has been shown that the shape of dark matter
haloes is mainly decided by the amount of anisotropy
in residual velocity distribution.   On the other
hand, the contribution of rotation has exhibited
only a minor effect on the meridional plane, and
no effect on the equatorial plane, as bifurcation
points occur for larger values of $\lambda$.   To
this respect, dark matter haloes
have been found to resemble giant elliptical 
galaxies (e.g., Binney 1976; Binney \& Tremaine 
1987, Chap.\,4, \S\,3).

\appendix
\section{Different normalizations for
homeoidally striated density profiles}
\label{norma}

With regard to homeoidally striated density
profiles, defined by Eqs.\,(\ref{seq:profg}),
the mass, the inertia tensor, and the
self-energy tensor, may be expressed as (CM03):
\begin{lefteqnarray}
\label{eq:Md}
&& M=\nu_{mas}^\dagger M^\dagger~~; \\
\label{eq:Ipqd}
&& I_{pq}=\delta_{pq}\nu_{inr}^\dagger M^\dagger
(a_p^\dagger)^2~~; \\
\label{eq:Espqd}
&& (E_{sel})_{pq}=-\nu_{sel}^\dagger\frac{G(M^
\dagger)^2}{a_1^\dagger}(B_{sel})_{pq}~~;
\end{lefteqnarray}
where the profile parameters, $\nu^\dagger$,
take the explicit form (CM03):
\begin{lefteqnarray}
\label{eq:numd}
&& \nu_{mas}^\dagger=\frac32\int_0^\Xi F(\xi)
\diff\xi~~; \\
\label{eq:nuid}
&& \nu_{inr}^\dagger=\frac32\int_0^\Xi F(\xi)
\xi^2\diff\xi~~; \\
\label{eq:nusd}
&& \nu_{sel}^\dagger=\frac9{16}\int_0^\Xi F^2(\xi)
\diff\xi~~;
\end{lefteqnarray}
and the function, $F(\xi)$, has the definition
(Roberts 1962):
\begin{equation}
\label{eq:Fd}
F(\xi)=2\int_\xi^\Xi f(\xi^\prime)\xi^\prime
\diff\xi^\prime~~;
\end{equation}
where the function, $f(\xi)$, corresponds to
the density profile, according to Eqs.\,(\ref
{seq:profg}).   It can easily be seen that the
following relations hold:
\begin{equation}
\label{eq:FdFd}
F(\Xi)=0~~;\qquad\frac{\diff F}{\diff\xi}=-2
\xi f(\xi)~~;
\end{equation}
and an integration by parts 
shows that:
\begin{equation}
\label{eq:Ff}
\int_0^\Xi f(\xi)\xi^n\diff\xi=\frac{n-1}2
\int_0^\Xi F(\xi)\xi^{n-2}\diff\xi~~;\qquad
n>1~~;
\end{equation}
which allows the calculation of the profile
parameters.   The scaling mass, $M^\dagger$,
has the same definition in CM03 as its
counterpart, $M_0$, in the current paper,
according to Eq.(\ref{eq:M0}).   Then the
following relation holds:
\begin{equation}
\label{eq:McM0}
M^\dagger=M_0~~;
\end{equation}
for further details, see CM03.

The mass, the inertia tensor, and the 
self-energy tensor, in the current paper are
expressed by Eqs.(\ref{eq:M}), (\ref{eq:Ipq}),
and (\ref{eq:Espq}), respectively.   The
comparison with their counterparts, expressed
by Eqs.\,(\ref{eq:Md}), (\ref{eq:Ipqd}), and 
(\ref{eq:Espqd}), yields:
\begin{lefteqnarray}
\label{eq:nunum}
&& \nu_{mas}=\nu_{mas}^\dagger~~; \\
\label{eq:nunui}
&& \nu_{inr}=\nu_{inr}^\dagger\frac{M^\dagger}M
\frac{(a_p^\dagger)^2}{a_p^2}~~; \\
\label{eq:nunus}
&& \nu_{sel}=\nu_{sel}^\dagger\left(\frac{M^
\dagger}M\right)^2\frac{a_1}{a_1^\dagger}~~;
\end{lefteqnarray}
where the last two relations, owing to
Eqs.\,(\ref{eq:profgb}), (\ref{eq:M}),
(\ref{eq:McM0}), and (\ref{eq:nunum}), may be 
written under the equivalent form:
\begin{lefteqnarray}
\label{eq:nuimc}
&& \nu_{inr}=\frac{\nu_{inr}^\dagger}{\nu_
{mas}^\dagger\Xi^2}~~; \\
\label{eq:nusmc}
&& \nu_{sel}=\frac{\nu_{sel}^\dagger\Xi}
{(\nu_{mas}^\dagger)^2}~~;
\end{lefteqnarray}
and the profile parameters, $\nu_{mas}^\dagger$,
$\nu_{inr}^\dagger$, $\nu_{sel}^\dagger$, are
expressible as functions of the scaled radius,
$\Xi$ (CM03).

The angular-momentum vector and the 
rotational-energy tensor, for the special
rotational velocity fields discussed in
Sect.\,\ref{gente}, may be expressed as
(CM03):
\begin{lefteqnarray}
\label{eq:Jsd}
&& J_s=\delta_{s3}\eta_{anm}^\dagger\nu_{anm}^
\dagger M^\dagger a_p^\dagger(1+\epsilon_{qp}^2)
(v_{rot}^\dagger)_p~~;\quad p\ne q\ne s~~; \\
\label{eq:Erpqd}
&& (E_{rot})_{pq}=\delta_{pq}(1-\delta_{p3})
\eta_{rot}^\dagger\nu_{rot}^\dagger M^\dagger
\left[(v_{rot}^\dagger)_p\right]^2~~;
\end{lefteqnarray}
and the related module (angular momentum)
and trace (rotational energy), respectively, as:
\begin{lefteqnarray}
\label{eq:Jd}
&& J=\eta_{anm}^\dagger\nu_{anm}^\dagger M^\dagger 
a_1^\dagger(1+\epsilon_{21}^2)(v_{rot}^\dagger)_1~~; \\
\label{eq:Erd}
&& E_{rot}=\eta_{rot}^\dagger\nu_{rot}^\dagger M^\dagger
(1+\epsilon_{21}^2)\left[(v_{rot}^\dagger)_1\right]^2~~;
\end{lefteqnarray}
where $(v_{rot}^\dagger)_p$ is the rotational
velocity at the top axis, $a_p^\dagger$, of 
the reference isopycnic surface.   

The profile parameters, $\nu^\dagger$, and 
the shape parameters, $\eta^\dagger$, take 
the explicit form (CM03):
\begin{lefteqnarray}
\label{eq:nujd}
&& \nu_{anm}^\dagger=\int_0^\Xi\frac{\Omega
(\xi,0)}{\Omega(1,0)}f(\xi)\xi^4\diff\xi~~; \\
\label{eq:nurd}
&& \nu_{rot}^\dagger=\int_0^\Xi\frac{\Omega^2
(\xi,0)}{\Omega^2(1,0)}f(\xi)\xi^4\diff\xi~~; \\
\label{eq:etajd}
&& \eta_{anm}^\dagger=\frac34\epsilon^4\int_{-\pi/2}
^{+\pi/2}\frac{\Omega(1,\theta)}{\Omega(1,0)}\frac
{\sin^3\theta\diff\theta}{(\cos^2\theta+
\epsilon^2\sin^2\theta)^{5/2}}~~; \\
\label{eq:etard}
&& \eta_{rot}^\dagger=\frac38\epsilon^4\int_{-\pi/2}
^{+\pi/2}\frac{\Omega^2(1,\theta)}{\Omega^2(1,0)}\frac
{\sin^3\theta\diff\theta}{(\cos^2\theta+
\epsilon^2\sin^2\theta)^{5/2}}~~;
\end{lefteqnarray}
where $\epsilon$ is the ratio of polar to
equatorial axis and $\theta$ is the azimuthal
angle.   It is worth of note that the above
expressions for the shape parameters, $\eta^
\dagger$, hold for axisymmetric configurations.
The shape parameters, $\eta$, are also expressed
by Eqs.\,(\ref{eq:etajd}) and (\ref{eq:etard}),
provided $\Omega(1,\theta)$ and $\Omega(1,0)$
therein are replaced by $\Omega(\Xi,\theta)$
and $\Omega(\Xi,0)$, respectively.   The
extension of the above definitions and results
to (necessarily) solid-body rotating, triaxial
configurations, yields $\eta_{anm}=1$, $\eta_
{rot}=1/2$, according to 
Eqs.\,(\ref{eq:Jpqrr})-(\ref{eq:Errr}).   For
further details, see CM03. 

The angular-momentum vector, the
rotational-energy tensor, the angular momentum,
and the rotational energy, in the current
paper are expressed by Eqs.\,(\ref{eq:Jpq}), 
(\ref{eq:Erpq}), and (\ref{eq:J}), (\ref{eq:Er}),
respectively.   The comparison with their
counterparts, Eqs.\,(\ref{eq:Jsd}), (\ref
{eq:Erpqd}), and (\ref{eq:Jd}), (\ref{eq:Erd}),
respectively, yields:
\begin{lefteqnarray}
\label{eq:nunuj}
&& \nu_{anm}=\nu_{anm}^\dagger\frac{\eta_{anm}^
\dagger}{\eta_{anm}}\frac{M^\dagger}M\frac{a_p^
\dagger}{a_p}\frac{(v_{rot}^\dagger)_p}{(v_{rot})
_p}~~;\qquad p=1,2~~; \\
\label{eq:nunur}
&& \nu_{rot}=\nu_{rot}^\dagger\frac{\eta_{rot}^
\dagger}{\eta_{rot}}\frac{M^\dagger}M\frac
{(v_{rot}^\dagger)_p^2}{(v_{rot})_p^2}~~;
\qquad p=1,2~~;
\end{lefteqnarray}
where, owing to Eq.\,(\ref{eq:rvang}), $\eta_
{anm}=\eta_{anm}^\dagger$ and $\eta_{rot}=
\eta_{rot}^\dagger$.   Then Eqs.\,(\ref
{eq:nunuj}) and (\ref{eq:nunur}), by use of
(\ref{eq:profgb}), (\ref{eq:M}), and
(\ref{eq:McM0}), take the equivalent form:
\begin{lefteqnarray}
\label{eq:nuamc}
&& \nu_{anm}=\frac{\nu_{anm}^\dagger}{\nu_
{mas}^\dagger\Xi}\frac{(v_{rot}^\dagger)_p}
{(v_{rot})_p}=\frac{\nu_{anm}^\dagger}{\nu_
{mas}^\dagger\Xi^2}\frac{\Omega(1,0)}
{\Omega(\Xi,0)}~~;\qquad p=1,2~~; \\
\label{eq:nurmc}
&& \nu_{rot}=\frac{\nu_{rot}^\dagger}
{\nu_{mas}^\dagger}\frac{(v_{rot}^\dagger)_
p^2}{(v_{rot})_p^2}=\frac{\nu_{rot}^\dagger}
{\nu_{mas}^\dagger\Xi^2}\frac{\Omega^2(1,0)}
{\Omega^2(\Xi,0)}~~;\qquad p=1,2~~;
\end{lefteqnarray}
where $(v_{rot}^\dagger)_p/(v_{rot})_p=1/\Xi$
if the system is in rigid rotation, and $(v_
{rot}^\dagger)_p/(v_{rot})_p=1$ if the 
rotational velocity is constant on the 
equatorial plane.
The profile parameters, $\nu_{anm}^\dagger$,
$\nu_{rot}^\dagger$, are expressible as
functions of the scaled radius, $\Xi$ (CM03).

The above results allow the calculation of the
profile parameters, $\nu$, appearing in the
text.   The corresponding, explicit expressions,
may be seen in Caimmi \& Marmo (2004).

\section{Limiting configurations}
\label{lico}

The general expressions of the shape factors, 
$A_p$, and related quantities, for limiting
configurations of ellipsoidal shape reduce to
undetermined expressions of the kind $0/0$, 
$\infty/\infty$, {\it et similia} (e.g.,
Caimmi 1991, 1992, 1995).  Then explicit
calculations are needed for round,
oblate, prolate, flat, and oblong configurations,
which can be found in the above quoted
references.   Though flat and oblong configurations
are dynamically unstable (e.g., Jeans 1929, 
Chap.\,IX), they shall also be included for sake
of completeness.
Accordingly, the expression
of the shape factors in the self-energy tensor,
$(B_{sel})_{pp}=\epsilon_{p2}\epsilon_{p3}A_p$, 
see Eq.\,(\ref{eq:Bspq}), must be determined
even for flat and oblong configurations.

The shape factors, $A_p$, depend on the axis 
ratios, via elliptic integrals of first and
second kind (e.g., Caimmi 1992):
\begin{leftsubeqnarray}
\slabel{eq:FEa}
&& F(\beta,p)=\int_0^\beta\frac{\diff\phi}{\sqrt 
{1-p^2\sin^2\phi}}~~;\qquad E(\beta,p)=\int_0^\beta
\sqrt{1-p^2\sin^2\phi}\diff\phi~~; \\
\slabel{eq:FEb}
&& \beta=\arccos\epsilon_{31}~~;\qquad p=\frac
{e_{21}}{e_{31}}~~;\qquad e_{pq}=\sqrt{1-\epsilon_
{pq}^2}~~; 
\label{seq:FE}
\end{leftsubeqnarray}
where $\beta=\pi/2$ for both flat and oblong 
configurations, while $p=e_{21}$ and $p=1$ 
for flat and oblong configurations, respectively.

With regard to the self-energy tensor, by use
of the general expression of the shape factors, 
$A_p$, derived in Caimmi (1992; therein defined
as $\hat{\alpha}_p$), the flat limit ($\epsilon
_{21}>0$, $\epsilon_{31}\to0$) reads:
\begin{leftsubeqnarray}
\slabel{eq:Bsfa}
&& (B_{sel})_{11}=\epsilon_{12}\epsilon_{13}A_1=
\frac2{e_{21}^2}\left[F\left(\frac\pi2,e_{21}
\right)-E\left(\frac\pi2,e_{21}\right)\right]
=b_1~~; \\
\slabel{eq:Bsfb}
&& (B_{sel})_{22}=\epsilon_{22}\epsilon_{23}A_2=
\frac2{e_{21}^2}\left[E\left(\frac\pi2,e_{21}
\right)-\epsilon_{21}^2F\left(\frac\pi2,e_{21}
\right)\right]=b_2~~; \\
\slabel{eq:Bsfc}
&& (B_{sel})_{33}=\epsilon_{32}\epsilon_{33}A_3=
\frac2{e_{21}^2}\left[-e_{21}^2E\left(\frac\pi2,
e_{21}\right)+\frac{\epsilon_{31}}{\epsilon_{21}}
e_{21}\right]=0~~;
\label{seq:Bsf}
\end{leftsubeqnarray}
and the special case of the oblate, flat limit
($\epsilon_{21}=1$, $\epsilon_{31}\to0$) reads: 
\begin{leftsubeqnarray}
\slabel{eq:Bsofa}
&& (B_{sel})_{11}=(B_{sel})_{22}=\lim_{\epsilon
_o\to0}\left(\epsilon_o^{-1}\alpha\right)=\frac
\pi2~~;\qquad(B_{sel})_{33}=\lim_{\epsilon
_o\to0}\left(\epsilon_o\gamma\right)=0~~; \\
\slabel{eq:Bsofb}
&& \epsilon_o=\epsilon_{31}=\epsilon_{32}~~;
\qquad\alpha=\lim_{\epsilon_{21}\to1}A_1=\lim
_{\epsilon_{21}\to1}A_2~~;\qquad\gamma=\lim_
{\epsilon_{21}\to1}A_3~~;
\label{seq:Bsof}
\end{leftsubeqnarray}
for further details see e.g., Caimmi (1991).

Oblong configurations may be conceived in a
twofold manner, according if $\epsilon_{21}
\to\epsilon_{31}=0$ (flat-oblong, hereafter 
quoted as f-oblong) or $\epsilon_{21}=
\epsilon_{31}\to0$ (prolate-oblong, hereafter
quoted as p-oblong).
In any case, $\beta\to\pi/2$, $p\to1$, $e_
{21}\to1$, and Eqs.\,(\ref{eq:FEa}) reduce to: 
\begin{equation}
\label{eq:FEo}
\lim_{\epsilon_{21}\to0}F\left(\frac\pi2,e_{21}
\right)=\lim_{x\to\pi/4}\ln\frac{1+\tan x 
}{1-\tan x}=+\infty~~;\qquad E\left(
\frac\pi2,1\right)=1~~;
\end{equation}
in addition, using the series development
of the complete elliptic integral of first
kind and Wallis' product (e.g., Spiegel 1968,
\S\S\,34.2 and 38.9) yields: 
%
%
\begin{equation}
\label{eq:lW}
\lim_{\epsilon_{21}\to0}\epsilon_{21}^2F\left(
\frac\pi2,e_{21}\right)=\frac\pi2\lim_{n\to+
\infty}\left[\frac{1\cdot3\cdot\dots\cdot(2n-1)}
{2\cdot4\cdot\dots\cdot2n}\right]^2=\lim_{n\to+
\infty}\frac1{2n+1}=0~~;
\end{equation}
with regard to f-oblong configurations, 
Eqs.\,(\ref{seq:Bsf}) reduce to:
\begin{equation}
\label{eq:Bfo}
(B_{sel})_{11}=+\infty~~;\qquad(B_{sel})_{22}=2~~;
\qquad(B_{sel})_{33}=0~~;
\end{equation}
on the other hand, with regard to p-oblong
configurations, the shape factors in the 
self-energy tensor take the expression:
\begin{leftsubeqnarray}
\slabel{eq:Bspfa}
&& (B_{sel})_{11}=\lim_{\epsilon_p\to0}\left(\epsilon_p^
{-2}\gamma\right)=\lim_{\epsilon\to+\infty}\epsilon^2
\frac2{\epsilon^2-1}\left[\epsilon\frac{\arcsinh\sqrt
{\epsilon^2-1}}{\sqrt{\epsilon^2-1}}-1\right]=+\infty
~~; \\
\slabel{eq:Bspfb}
&& (B_{sel})_{22}=(B_{sel})_{33}=\lim_{\epsilon_p\to0}
\alpha=\lim_{\epsilon\to+\infty}\frac\epsilon{\epsilon
^2-1}\left[\epsilon-\frac{\arcsinh\sqrt{\epsilon^2-1}}
{\sqrt{\epsilon^2-1}}\right]=1~~; \\
\slabel{eq:Bspfc}
&& \epsilon_p=\epsilon^{-1}=\epsilon_{31}=\epsilon_{21}
~~;\qquad\alpha=\lim_{\epsilon_{21}\to\epsilon_{31}}A_1
~~;\qquad\gamma=\lim_{\epsilon_{21}\to\epsilon_{31}}A_2
=\lim_{\epsilon_{21}\to\epsilon_{31}}A_3~~;
\label{seq:Bspf}
\end{leftsubeqnarray}
for further details see Caimmi (1991).

The comparison between Eqs.\,(\ref{eq:Bfo}) and
(\ref{seq:Bspf}) discloses that the shape 
factors in the self-energy tensor exhibit
a discontinuity passing from f-oblong to
p-oblong configurations (e.g., Caimmi 1993b).   
The values of the axis ratios, $\epsilon_{21}$
and $\epsilon_{31}$, the shape factors,
$A_p$ and $(B_{sel})_{pp}$, related to
limiting configurations of ellipsoidal 
shape, are listed in Tab.\,\ref{t:lico}
where the upper part, containing the 
shape factors, $A_p$, has been taken from
Caimmi (1995).
\begin{table}
\begin{tabular}{lccccccc}
\hline
\hline
\multicolumn{1}{c}{shape factor} &
\multicolumn{6}{c}{limiting configuration} \\
\multicolumn{1}{c}{} &
\multicolumn{1}{c}{p-oblong} &
\multicolumn{1}{c}{f-oblong} &
\multicolumn{1}{c}{flat} &
\multicolumn{1}{c}{f-oblate} &  
\multicolumn{1}{c}{prolate} &
\multicolumn{1}{c}{oblate} &
\multicolumn{1}{c}{round} \\
\hline
$\epsilon_{21}$ & \phantom{}0 & \phantom{}0 & $\epsilon_f$ & 1 &
$\epsilon_p$ & 1 & 1 \\
$\epsilon_{31}$ & 0 & \phantom{}0 & 0 & 0 & $\epsilon_p$ & 
$\epsilon_o$ & 1 \\
$A_1$ & \phantom{}0 & \phantom{}0 & 0 & 0 & $\gamma$ & $\alpha$ & 2/3 \\
$A_2$ & \phantom{}1 & \phantom{}0 & 0 & 0 & $\alpha$ & $\alpha$ & 2/3 \\
$A_3$ & \phantom{}1 & \phantom{}2 & 2 & 2 & $\alpha$ & $\gamma$ & 2/3 \\
$(B_{sel})_{11}$ & \phantom{}$+\infty$ & \phantom{}$+\infty$ & $b_1$ & 
$\pi/2$ & $\epsilon_p^{-2}\gamma$ & $\epsilon_o^{-1}\alpha$ & 2/3 \\
$(B_{sel})_{22}$ & 1 & 2 & $b_2$ & $\pi/2$ & $\alpha$ & $\epsilon_o^{-1}
\alpha$ & 2/3 \\
$(B_{sel})_{33}$ & 1 & 0 & 0 & 0 & $\alpha$ & $\epsilon_o\gamma$ & 2/3 \\
\hline\hline
\end{tabular}
\caption{Values of shape factors, $A_p$ and 
$(B_{sel})_{pp}$, related to limiting
configurations defined by the values of the
axis ratios, $\epsilon_{21}$ and $\epsilon_
{31}$.} 
\label{t:lico}
\end{table}
The range of validity of the independent
variables, $\epsilon_f$, $\epsilon_p$,
$\epsilon_o$, and $\epsilon$, and the definition
of the functions, $\alpha$, $\gamma$, are:
\begin{leftsubeqnarray}
\slabel{eq:doepa}
&& 0<\epsilon_f\le1~~;\qquad0<\epsilon_p<1~~;\qquad
0<\epsilon_o<1~~; \\
\slabel{eq:doepb}
&& 0<\epsilon<1~,~~{\rm oblate;}\qquad
1<\epsilon<+\infty~,~~{\rm prolate;}
\label{seq:doep}
\end{leftsubeqnarray}
\begin{leftsubeqnarray}
\slabel{eq:algaa}
&& \alpha=
\cases{\displayfrac{\epsilon}{1-\epsilon^2}\left[
\displayfrac
{\arcsin(1-\epsilon^2)^{1/2}}{(1-\epsilon^2)^
{1/2}}-\epsilon\right]~~; & oblate; \cr
& \cr
\displayfrac{\epsilon}{\epsilon^2-1}\left[\epsilon-
\displayfrac
{\arcsinh(\epsilon^2-1)^{1/2}}{(\epsilon^2-1)^
{1/2}}\right]~~; & prolate; \cr} \\
&& \nonumber \\
&& \nonumber \\
\slabel{eq:algab}
&& \gamma=
\cases{\displayfrac2{1-\epsilon^2}\left[1-\epsilon
\displayfrac
{\arcsin(1-\epsilon^2)^{1/2}}{(1-\epsilon^2)^
{1/2}}\right]~~; & oblate; \cr
& \cr
\displayfrac2{\epsilon^2-1}\left[\epsilon
\displayfrac
{\arcsinh(\epsilon^2-1)^{1/2}}{(\epsilon^2-1)^
{1/2}}-1\right]~~; & prolate; \cr}
\label{seq:alga}
\end{leftsubeqnarray}
and the functions, $b_1$ and $b_2$, are expressed
by Eqs.\,(\ref{eq:Bsfa}) and (\ref{eq:Bsfb}).   For
further details, see Caimmi (1995).

\section{Expression of anisotropy and rotation
parameters in compact notation}
\label{foco}

The effective anisotropy parameters, $\tilde
{\zeta}_{pp}$, expressed by Eqs.\,(\ref{eq:zq}),
and the rotation parameters, $h$,
${\cal E}_{rot}$, $\upsilon$, $\lambda$, 
and $\chi_v$, expressed
by Eqs.\,(\ref{eq:hz}), (\ref{eq:Erz}),
(\ref{eq:upz}), (\ref{eq:lam3}), and 
(\ref{eq:vrope}), respectively, 
may be written in compact notation, using 
Eqs.\,(\ref{eq:Spq}) and (\ref{eq:Rpq}).   
The result is:
\begin{lefteqnarray}
\label{eq:czq}
&& \tilde{\zeta}_{pp}=\frac{\zeta_{pp}}\zeta=
\frac{{\cal S}_{pp}}
{{\cal S}}\left[1-2h\frac{{\cal R}_{pp}}
{{\cal S}_{pp}}\right]\left[1-2h\frac{{\cal
R}}{{\cal S}}\right]^{-1}~~; \qquad p=1,2,3~~; 
\end{lefteqnarray}
with regard to anisotropy parameters;
\begin{lefteqnarray}
\label{eq:ch}
&& h=\frac12\frac{{\cal S}}{{\cal R}}\left[
1-\frac\zeta{\zeta_{33}}\frac{{\cal S}_{33}}
{{\cal S}}\right]~~; \\
\label{eq:crErs}
&& {\cal E}_{rot}=\frac{{\cal R}}{{\cal S}} 
h~~; \\
\label{eq:cupsh}
&& \upsilon=\frac13\frac1{\eta_{rot}\nu_{rot}}
\frac{\epsilon_{21}\epsilon_{31}}{1+\epsilon_
{21}^2}{\cal S}\left[1-\frac{\zeta}{\zeta_{33}}
\frac{{\cal S}_{33}}{{\cal S}}\right]~~; \\
\label{eq:clam3}
&& \lambda^2=\frac14\frac{{\cal S}^2}
{{\cal R}}\left[1-\frac\zeta{\zeta_{33}}\frac
{{\cal S}_{33}}{{\cal S}}\right]\left[
1-\frac{1-\zeta}\zeta\frac\zeta{\zeta_{33}}
\frac{{\cal S}_{33}}{{\cal S}}\right]~~; \\
\label{eq:cvrope}
&& \chi_v^2=\zeta\left[\frac{\zeta_{33}}
\zeta\frac{{\cal S}}{{\cal S}_{33}}-1
\right]~~;
\end{lefteqnarray}
with regard to rotation parameters.

Finally, the combination of Eqs.\,(\ref{eq:ch}),    
(\ref{eq:cupsh}), (\ref{eq:clam3}), and (\ref
{eq:cvrope}) yields:
\begin{lefteqnarray}
\label{eq:cupfh}
&& \upsilon=\frac23\frac1{\eta_{rot}\nu_{rot}}
\frac{\epsilon_{21}\epsilon_{31}}
{1+\epsilon_{21}^2}{\cal R}h~~; \\
\label{eq:clam4}
&& \lambda^2=\frac12{\cal S}h\left[
1-\frac{1-\zeta}\zeta\frac\zeta{\zeta_{33}}
\frac{{\cal S}_{33}}{{\cal S}}\right]~~; \\
\label{eq:cvrofe}
&& 
\chi_v^2=\zeta\frac{2{\cal R}h}{{\cal S}-2
{\cal R}h}~~;
\end{lefteqnarray}  
which allows a connection between different
rotation parameters.

An inspection of Eq.\,(\ref{eq:czq}) shows that the
rotation parameter, $h$, attains the special values:
\begin{leftsubeqnarray}
\slabel{eq:zitha}
&& h=\frac12\frac{{\cal S}_{qq}}{{\cal R}_{qq}}~~;
\qquad\tilde{\zeta}_{qq}=0~~;\qquad q=1,2~~; \\
\slabel{eq:zithb}
&& h=\frac12\frac{{\cal S}-{\cal S}_{pp}}{{\cal R}-
{\cal R}_{pp}}~~;\qquad\tilde{\zeta}_{pp}=1~~;\qquad
p=1,2,3~~; \\
\slabel{eq:zithc}
&& h=0~~;\qquad\tilde{\zeta}_{pp}=\frac{{\cal S}_{pp}}
{{\cal S}}~~;\qquad p=1,2,3~~;
\label{seq:zithc}
\end{leftsubeqnarray}  
where the last relation is equivalent to Eq.\,(\ref
{eq:zp}).


\end{document}